\documentclass[11pt]{article}

\usepackage[utf8]{inputenc} 

\usepackage{lscape} % or \usepackage{pdflscape} for PDF output
\usepackage{algorithm}
\usepackage{algpseudocode}
\usepackage{graphicx} 

\usepackage{mathtools}  
\usepackage[table,xcdraw]{xcolor}
\usepackage{float}   % floatus
\usepackage{tabularx}              % more table options
\usepackage[l2tabu,orthodox]{nag}  % force newer (and safer) LaTeX commands
\usepackage{titling}               % allow redefinition of title formatting
\usepackage{imakeidx}              % create and index of words
\usepackage{xcolor}                % more colour options
\usepackage{enumitem}              % more list formatting options
\usepackage{amsmath, amssymb}
\usepackage{sectsty}
\usepackage[affil-it]{authblk}
\usepackage{hyperref}
\usepackage{multirow}
\usepackage{booktabs} % for much better looking tables
\usepackage{array} % for better arrays (eg matrices) in maths
\usepackage{verbatim} % adds environment for commenting out blocks of text & for better verbatim
\usepackage{subfig} % make it possible to include more than one captioned figure/table in a single float
\usepackage{multirow}
\hypersetup{
    colorlinks,
    linkcolor={red!50!black},
    citecolor={blue!50!black},
    urlcolor={blue!80!black}
}
\usepackage[margin=3cm]{geometry}  % Adjust the margin as desired
\usepackage{fancyhdr} 
\allsectionsfont{\sffamily\mdseries\upshape} 

\usepackage[nottoc,notlof,notlot]{tocbibind} % Put the bibliography in the ToC
\usepackage[titles,subfigure]{tocloft} % Alter the style of the Table of Contents

\title{dYdX: Liquidity Providers' Incentive Programme Review}
\author{Colin Chan}
\date{June 2023}
\begin{document}
\maketitle

\begin{abstract}
\centering
Liquidity providers are currently incentivised to provide liquidity through the LP Incentives Programme. Based on the various parameters - makerVolume, depths and spreads, they are rewarded accordingly based on their activities. Given the maturity of the BTC and ETH markets, alongside other altcoins which enjoys a consistent amount of liquidity, this paper aims to update the formula to encourage more active and efficient liquidity, improving the overall trading experience. In this research, I begin by providing a basic understanding of spread management, before introducing the methodology with the various metrics and conditions. This includes gathering orderbooks on a minute interval and reconstructing the depths based on historical trades to establish an upper bound. I end off by providing recommendations to update the maxSpread parameter and alternative mechanisms/solutions to improve the existing market structures. \footnote{This research was sponsored by the dYdX Grants Programme. Thank you to the dYdX Trading Team, Foundation and Grants Team for the opportunity and their support. }
\newpage

\end{abstract}

\tableofcontents
\newpage
\section{Introduction}

Exchanges provide a medium for buyers and sellers to come together and trade at a common price based on each agents' perspective. With the advent of electronic trading, we witness the emergence of liquidity providers (LPs) who provide transparent two-sided markets at all times. They play a critical role to ensure a healthy flow of liquidity for the price discovery process and enable a seamless trading experience. \\

Through the years, a primary aim of exchanges is to attract and reward these LPs. The objective is to establish a mechanism to encourage a consistent of liquidity through rebates and incentives. For instance, NYSE establishes designated Market Makers (DMMs) where they are specially appointed and obliged to maintain quotes and facilitate transactions for an allocated set of stocks. Deutsche Boerse has also implemented the Xetra Liquidity Provider Programme to reward LPs based on a set of parameters. Especially for margin trading which relies more on liquidity depth to absorb delinquent traders, exchanges have to find a balance between tighter spreads and deeper depths. As such, to incentivise LPs, it is critical to understand their considerations in quoting buy and sell orders to the book (BIS, 2015): \\
\par
\par
Revenue: Bid ask spread, Delta market value from position 
\par
Costs: Funding,  Maker fees \\
\par

They primarily profit by arbitraging the differences in spread and thus compete on the order execution and turnover to build more volume. In fact, spread behaviour is a crucial focus for all market participants: 
\par
* Takers: Tighter spreads improve the price discovery mechanism as takers can execute orders closer to the fair price and institutional investors can put on larger blocks of trade without suffering from excessive slippage. 
\par
* Makers (LPs): These providers constantly face a dilemma between turnover and spreads as many attempt to identify the optimal spread management. For instance, lower spread means that the LPs’ quotes are more likely to get hit, resulting in higher turnover rates. However, this will also entail a lower income per trading cycle since the profit potential is lower each time. \\

Therefore, the paper will begin by unravelling the LP rewards scheme on dYdX, in particular the spread management aspects. Rather than just focussing on historical order book data and depths, I will empirically analyse the trading data via the dYdX API. In particular, this will seek to understand the exact volume of trades happening on the volume and provide more granular insights into the order book resiliency, especially after liquidity shocks initiated by anomalous market orders. The aim of this paper would be to: \\

\emph{Enhance the market efficiency by ensuring sufficient liquidity in the orderbook and tightening the maxSpread parameter, while being aligned with current liquidity provision strategies by LPs and trading statistics on the exchange}\\

\section{Deconstructing the Spread}

The bid-ask spread (quoted spread) can be defined as: 
\begin{equation}
	 Quoted Spread =Lowest Ask- Highest Bid
\end{equation}

This is the simplest form of measurement taken directly from quotes at the top of the orderbook. However, the majority of trades usually occur beyond this point given varying transaction sizes. \\

Prior empirical research by Demsetz (1968)  provides the foundation for identifying the determinants of the spread based on the NYSE - inventory, transaction, information costs and level of competition based on the number of LPs providing liquidity. This was further elaborated by Copeland and Galai (1983) in emphasising the variables of volatility and trading activities which are extrinsic risks faced by liquidity providers. More recent studies on the NYSE have also revealed that volume exerts statistically significant inverse relationship with spreads while the impact of price on spreads returns mixed results depending on the sectors as it incurs different variations in the market making costs. Furthermore, greater volatility induces more risks for LPs, who may widen spreads to price in the uncertainties. As a result, different exchanges have curated LP Programmes which actively encourage volume and spreads for designated LPs. For instance, the Xetra Liquidity Provider service agreement includes volatility within its scores, where the achieved best bid-offer (BBO) presence is doubled in times of volatility, on which the EUR index exceeds a range of 3\%. \\

Therefore, the spread can be represented with a simplified equation where:\\

\begin{equation}
	 Spread = f(l , v , p) 
\end{equation}\\
 l = liquidity,  v = volatility, p = price impact \\

It is commonly expected that the long tail markets will have a wider spread, reflecting greater adverse selection. Hence, compensating LPs for assuming delta risk in being exposed to these price movements.

\section {Understanding DYDX LP Rewards Incentives}

To incentivize two-sided liquidity, the LP rewards programme was introduced to distribute \$DYDX to LPs - makerVolume, depth and spread vs mid market, and uptime on provision. The general formula can be summarised with the participant’s Q score per epoch: 
\begin{equation}
Q_{\text{Final}} = \left[Q_{\text{min}} (\sum_{i=1}^{n} \frac{\text{BidDepth}_{i}}{\text{Spread}{_i}} + \sum_{j=1}^{m}\frac{\text{AskDepth}_{j}}{\text{Spread}_{j}}) \right]^y \times \left[\left(\sum_{} {{Q_{\text{min}} (N) > 0}}\right)\right]^5 \times \text{makerVolume}^z
\end{equation}

where:

$y = 0.15$ if BTC/ETH, $y = 0.35$ for the remaining. 
\par$z = 0.85$ for BTC/ETH, $z = 0.65$ for the remaining.\\

Previous community discussions have centred around adjusting the weights for the overall formula. For instance, LPs who generated lower makerVolumes received the majority of the LP rewards which contradicted the purpose of this programme.

\subsection{Different Parameters - Types of Liquidity Needed}
In the formula, the paper will expound deeper into the scoring mechanism based on Depth/Spread located within. The programme has specified a MaxSpread parameter where providers will have to provide bids and asks within this spread for the markets in order to qualify for the rewards.\\

\par Based on the formula, there are 2 segments that focus on spread management - depth/spread ratios and Qmin, while a weightage is placed on makerVolume. These parameters represent different interests from LPs: \\
\\
\par * Deep Liquidity: LPs who constantly provide 2 sided liquidity within the maxSpread will be eligible for these rewards, even though some of these remain as passive orders in the book. This influences their uptime scores which is raised to the power of 5.  
\par \begin{table}[h]
\centering
\caption{MaxSpread parameter}
\begin{tabular}{|c|c|cl}
\hline
Markets & Max Spread vs Mid Market \\
\hline
BTC, ETH & 20 bps \\
\hline
Remaining & 40 bps \\
\hline
\end{tabular}
\end{table}
* Active Liquidity: makerVolume refers to the amount of liquidity provided by LPs which were successfully taken by counteracting trades. These 'active liquidities' have a higher probability of being near the BBO, away from the maxSpread, providing a favourable environment for takers. \\

Thus, a key focus is to incentivise LPs to provide active and deep liquidity at competitive prices. \\

\subsection{Peer Analysis}

\subsubsection{CEXs' LP Programmes}

CEXs such as Binance and Okx implement LP programmes, which primarily incentivises LPs to trade through lower to negative maker fees. The main criteria includes the amount of assets held on the exchange, monthly trading volume and/or amount of CEX native tokens. For Binance, the exchange also implements weekly performance reviews, looking at makerVolumes, Bid/offer spreads, total order sizes, duration and market making time. \\

In traditional exchanges such as the NASDAQ, the Designated Liquidity Provider (DLP) Programme was implemented, aimed to encourage LPs to support a broader range of ETPs. This is mainly through tiered rebates (i.e. negative maker fees). Certain applicable criteria include time @NBBO (national best bid offer, across exchanges), time within 5bps of NBBO, notional depth and average spread. In particular, the focus is on ETFs that 'are under the 250,000 share threshold for average daily volume', where the exchange gives LPs a stipend on top of the existing rebates. These have yielded positive results with tighter spreads and depths around the NBBO. 

\subsubsection{Orderbook Snapshot}

As a comparison across exchanges, a  snapshot of the orderbooks on dYdX and the most liquid exchange, Binance, were taken on May 4 across all the markets supported using USDT as the base asset. UMA-USDT is not supported on Binance.   \\

It can be seen that the BBO spreads for certain markets such as ZRX-USD and UMA-USD have been $consistently \geq 40bps$ and likely suggests that LPs are not incentivised to provide liquidity in these markets. This observation is supported by the constant readjustments in margin fractions due to the lack of liquidity and volatility experienced.  \\

\begin{figure}[htbp]
    \centering
     \caption{BBO Spreads on dYdX vs Binance (where bid spread = ask spread)}

    \includegraphics[width=0.8\textwidth]{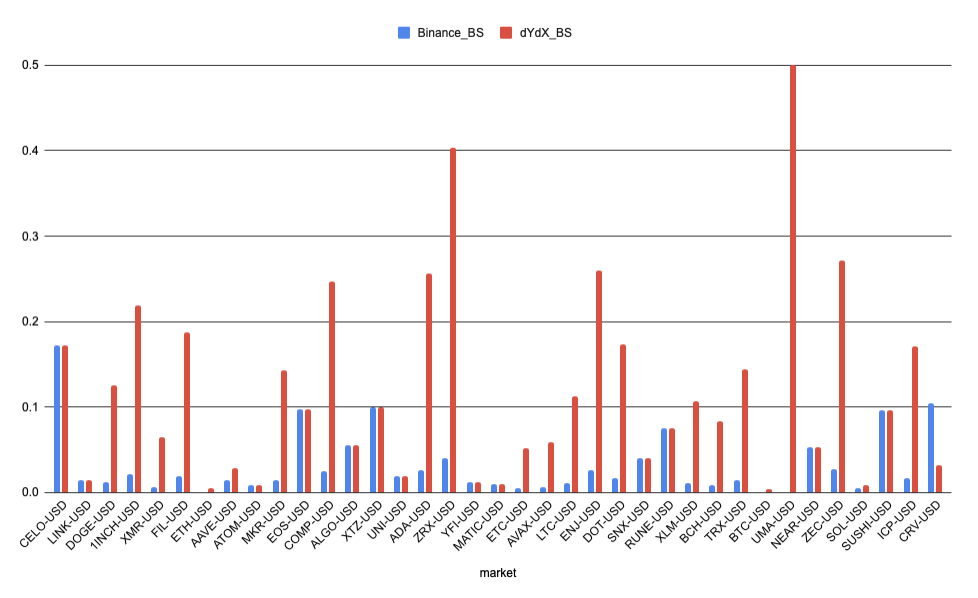}
    \end{figure}

A closer look at the liquidity distribution in the order book highlights the positioning of the limit orders. This is analogous to a competitive (non-cooperative) game where LPs aim to minimize the spreads to attract trading volume and increase the likelihood of executing trades based on the probability density function for the different price levels. Ultimately, a LP seeks to position themselves around the tightest spreads possible, thereby influencing other LPs to compete around the BBO spreads to have higher execution rate. The following shows a snapshot of the order books for BTC:

\begin{table}[h]
\centering
\caption{BTC-USD on dYdX}
\begin{tabular}{|c|c|}
\hline
Original maxSpread = 20bps & maxSpread = 10bps \\
\hline
\includegraphics[width=0.4\linewidth]{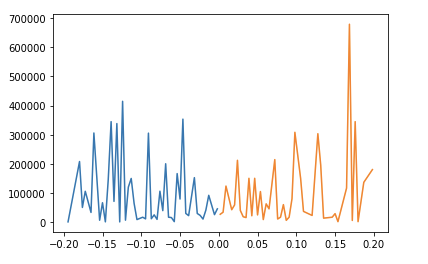} & \includegraphics[width=0.4\linewidth]{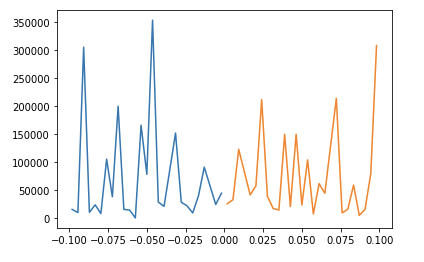} \\

\hline
\end{tabular}
\end{table}

\begin{table}[h]
\centering
\caption{BTC-USDT on Binance}
\begin{tabular}{|c|c|}
\hline
Original maxSpread = 20bps & maxSpread = 10bps \\
\hline
\includegraphics[width=0.4\linewidth]{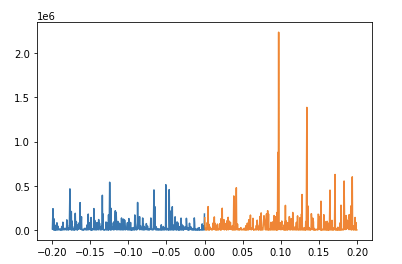} & \includegraphics[width=0.4\linewidth]{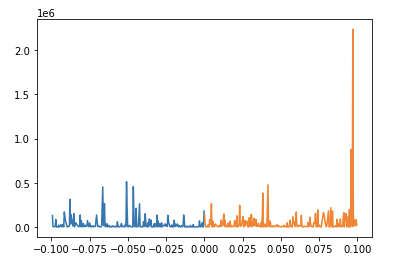} \\

\hline
\end{tabular}
\end{table}

\par From an absolute volume perspective, the majority of the volumes are found within 10bps from the midPrice for BTC-USD market. In this case, passive orders far away from the BBOs have a lower probability to get hit but yet LPs still continue to be eligible for these rewards should the depths hit the minimum requirements within the maxSpread. \\

Therefore, in determining the maxSpread parameter, it is crucial to understand the order book dynamics and contextualise it with the trades on the exchange. For instance, if majority of trades are low in volume, then perhaps incentivising too wide a spread may be unncessary.

\section{Methodology}

\subsection{Data Collection}

\subsubsection{Orderbook Data}

\par Minute by minute orderbook data was collected from the dYdX API for 2 weeks from May 11 (CPI Release) to May 24. This period was deliberately chosen given the harsh regulatory and economic climate (e.g. significant pullback of liquidity on CEXs). Specific events were also identified in which the orders removed a significant amount of liquidity from the market:\\

\begin{table}[h]
\centering
\small % Reduce font size
\setlength{\tabcolsep}{2pt} % Adjust column spacing
\caption{High Market Impact Events}
\begin{tabular}{|l|c|c|c|}
\hline
Date & Event \\
\hline
PPI Release &  Start : 1683804600,  End: 1683811800    \\\hline
US initial Jobless Claims & Start : 1684409400,  End: 1684416600 \\\hline
FOMC Meeting Minutes Release & Start : 1684947600, End : 1684954800 \\\hline
\end{tabular}
\end{table}

The order book data was manually collected on a minute by minute basis to determine the depths within the range of 10bps and 50bps.  I note that the data provided by the API does not distinguish between the different types of orders and hence, trade data was also retrieved to supplement the takers' behaviors .The following notation is used for the data within the order book at time \(t\): 

*\(m_t =  \frac{P_{bid, 0} + P_{ask, 0}}{2}\) --- mid price using the lowest ask and highest bid prices
\par
* \(p_{x,l}, v_{x,l}\) --- depicting the price and volume respectively for the \(l\)th level in the book, and the side (x = bid, ask)
\par
* \(ms_{x} = \text{midSpread}\) --- where bidSpread = askSpread in a symmetric book
\par
* \(liquidity_{pre}, liquidity_{post}\) --- depicts the amount of liquidity present before and after the market impact events

\subsubsection{Trade Data}

All trades executed between April 30 and May 24 were retrieved as well from the dYdX API. This contained the side, size, price and liquidation parameters within each trade executed. Subsequently, this will be grouped at the minute level to estimate the bid and ask depths necessary, effectively reconstructing the order book to suggest the tightest spreads possible based on this empirical analysis.

\subsubsection{Data Interpretation Measures}

Descriptive statistics are provided regarding the state of liquidity, to  understand the changes in order book depths and if there was sufficient buffer to absorb these trades. Furthermore, this will also summarize the daily trade statistics for each market, by identifying the following -  number of trades above \$50,000, \$100,000, \$250,000, \$500,000 and \$1.5m, mean, standard deviation and maximum trade size. \\
\par In particular, an analysis of LP behavior in highly stressed environments is critical to provide an initial understanding into the order book dynamics during these circumstances. For instance, LPs will exit markets when the risks of liquidity provision outweigh the benefits, and hence, the incentives scheme is necessary to retain LPs regardless of volatility. The resiliency analysis ( t - 1, t + 1 hours of Market Impact events) based on the identified events will visualise the changes in order book depths before and after the interval to understand the changes in volumes and how quick LPs were in enabling the order book to recover back to average levels. \\

\subsection{Metrics}

\subsubsection{Average Trade Size, Max Trade Size, Number of Trades above a certain amount on a daily basis}

\par

This identifies the amount of liquidity needed by takers to execute trades in a low slippage environment. Sufficient liquidity buffer will have to be available to fill these orders in an orderly fashion.

\subsubsection{Time to Recovery} 

\par For these Market Impact Events, a more granular perspective is adopted to fully understand the LPs' reaction. This will contribute to the general understanding in the order book resiliency under such stress tests as it is crucial to have a good liquidity recovery process in these instances. For a general perspective, these were calculated in minutes  to feature the duration of time before the amount of liquidity returns to 75\% of the average level. The metric was adapted from a study that emphasised the importance of resiliency as a market attribute in the Australian equity market. (Lo and Hall, 2015)

\subsubsection{(Upper Bound) : Estimated Depth Required}

Based on the trade data for each market, the estimated depth required will be calculated on each side of the order book. This is based on the reconstruction, aggregated per minute level, creating an upper bound. \\

\includegraphics[width=1\textwidth]{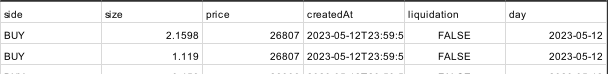}\\

In the BTC-USD market, if a trade with 'BUY" was recorded, the notional value of the trade (i.e. price *  size) is calculated. The trade time will be rounded off to the nearest minute based on 'createdAt'. This process is repeated for all trades. The trades are then grouped by hour and minute for each day, and added to the order book. In this case, it will be added to the 'Ask' side. Through this iterative approach, it creates an initial set of parameters for backtesting, to determine the upper bound of the recommended depths. A sample of the algorithm can be found below, which continuously updates the local state of the order book defined by the OrderBook class.\\
 \\

\begin{algorithm}[H]
  \caption{Reconstruction of Historical Order Book (Aggregated Per Minute)}
  \begin{algorithmic}[1]
  
\State $Global: \text{order\_books}, \text{times}$\\

\Procedure{Main}{Updating Orderbooks - Consolidates bids and asks}
    \For{$\text{\_}$ , $\text{trades\_minute}$ \textbf{in} $\text{trades\_df.groupby([trades\_df['hour'], trades\_df['min']])}$}
      \State $\text{trades\_minute} \gets \text{trades\_minute[['price', 'size', 'side']].values.tolist()}$
      \State $\text{order\_book} \gets \text{OrderBook()}$
      \State $\text{order\_book.reconstruct\_order\_book}(\text{trades\_minute})$
      \State $\text{times.append}(\_)$
      \State $\text{order\_books.append}(\text{order\_book})$
    \EndFor
\EndProcedure \\

\Function{Helper Class}{Orderbook}
\State \quad \textbf{def} reconstruct\_order\_book(self, trades):
    \State \quad \quad \textbf{self.bids} $\gets \{\}$
    \State \quad \quad \textbf{self.asks} $\gets \{\}$

\EndFunction

  \end{algorithmic}
 \end{algorithm}

\subsection{Conditions}

In deciding the maxSpread parameters, the following conditions will need to be fulfilled: 

\subsubsection{Order book and Trade Dynamics}

\par \textbf{Condition 1 (Lower Bound)} : $Min Tick Size$  \textgreater   $ Price \times Spread  $\\
\par For instance, ETC has a tick size of 0.01. At a price of \$18.33, should an LP quote at \$18.34, this is already equivalent to 5.5bps. This plays a critical role in influencing how LPs quote in the order book. (Refer to Appendix for Bps requirements). Therefore, if an LP who is quoting for UMA already incurred 42.8bps for just a 1 tick difference, then it will remain in the 40bps maxSpread basket. This represents a suggested lower bound for the market to remain attractive to LPs.\\

\par \textbf{Condition 2} : $Mean_{depth at chosen spread}$  $\geq$  $Mean_{volume at each side}(Per minute)$\\
 
Using historical trade data, the average size for each side of the order book is then compared against the depths retrieved in intervals of 10bps. The reconstructed depths will have to be greater than the 95th percentile of daily historical volumes, recorded on a minute by minute basis.  \\
\par Based on the isolated periods of volatility, the depths at the chosen spread will then be plotted against the trade volume recorded for that minute. This stress test will reveal if there's empirically sufficient liquidity buffer in Section 5.2. \\

\textbf{Condition 3} : Depth at chosen spread exhibits mean reverting behavior\\

\par The actual depth based on the chosen spread that has fulfilled Conditions 1 and 2 is chosen and this undergoes the statistical ADF test if the series is stationary - mean, variance, covariance and standard deviation are not a function of time, without trend or seasonal components. This will show that these liquidity depths are consistent with present LP behaviors and aren't mere coincidences. \\

\par \textbf{Condition 4} : $Depth_{spread, x}$  $\geq$  $Estimated_Volume_{x, t}(Per minute)$, where x = bid, ask\\

\par This is based on the event study for high market impact events in the later section.

\section{Results}

\subsection{Descriptive Statistics}

\subsubsection{Depth and Spread (Altcoins)}

Based on the reconstruction of the orderbook, aggregated per minute level, there is a distinct segregation amongst the different markets. During the recorded periods where large trade sizes are executed, \\ 
\par 

-- SOL leads the pack, requiring the largest upper bound of between \$300,000 and \$680,000. \\

-- AVAX, DOGE, MATIC, LINK, LTC, FIL, ATOM are in the next bracket, requiring a relatively larger amount of depths in the  support trading activities, \$150,000 up to \$300,000. \\

-- Next, 1INCH, AAVE, ADA, BCH, CRV, EOS, ICP, SUSHI, SNX, XTZ, COMP, MKR saw an estimated depth of under \$150,000, while the remaining markets require depths of less than \$100,000 and some lower than \$50,000. \\

However, it should be recognised that this means on the majority of occasions, the exchange may possibly be overtly incentivising for liquidity provision on normal days. This can be seen from the 95th percentile of bid and ask depths required in Appendix 10.3 and 10.4. \\ 

** Note that the mean is based on the number of minutes where trades are recorded.

\subsubsection{Trade Statistics}

By observing individual trades, the number of trades that have a size of over \$50,000, \$100,000, \$250,000, \$500,000, \$1.5 million are retrieved for each market, alongside the mean, standard deviation and maximum trades. The following observations can be noted: \\

Between April 30 and May 19: \\
\par 
-- Markets hardly recorded any trades with a notional value of above \$50,000, with the exception of SOL which had 44 trades of this size. 

-- By narrowing down on those with trades above \$100,000, the following markets are observed with the maximum recorded of 6 in the ATOM-USD market on May 8. \\ 

\begin{table}[h]
\centering
\caption{Number of trades above the amounts - \$50k and \$100k}
\begin{tabular}{|c|c|}
\hline
Trades above \$50,000 & Trades above \$100,000\\
\hline
\includegraphics[width=0.5\linewidth]{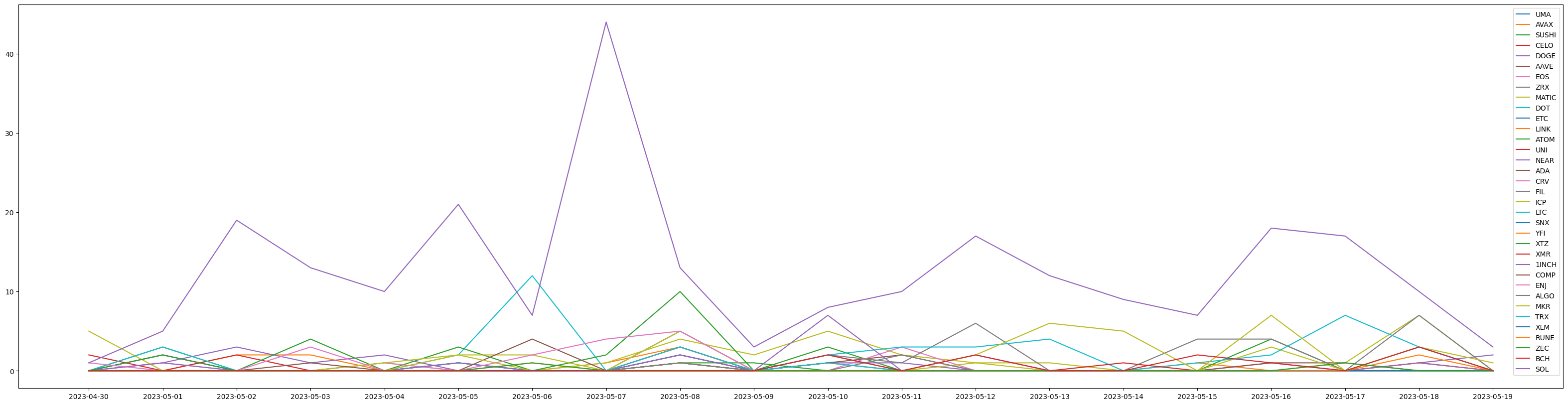} & \includegraphics[width=0.5\linewidth]{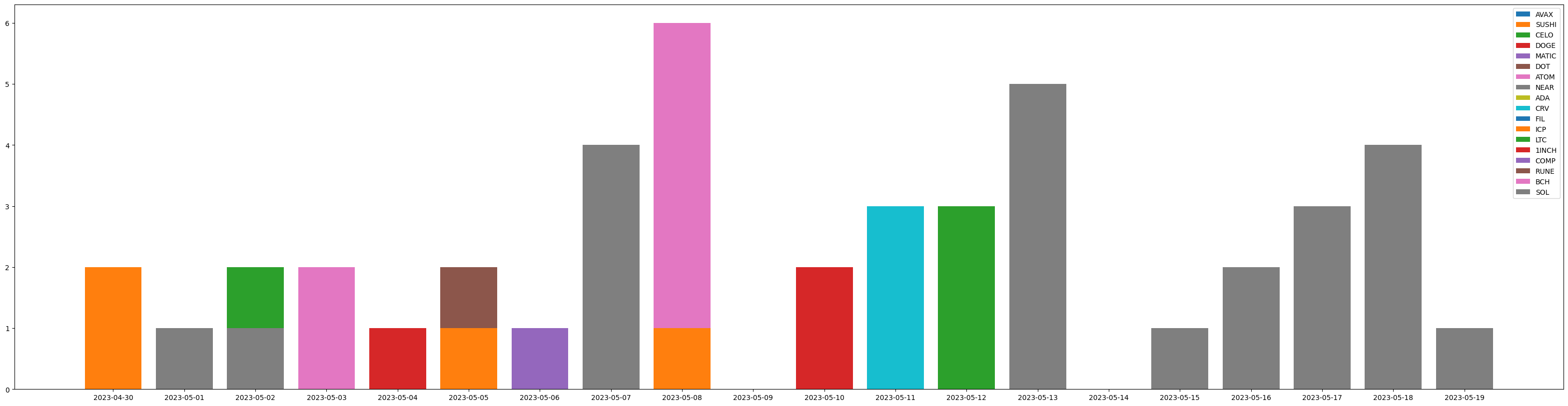} \\

\hline
\end{tabular}
\end{table}

Between May 20 and May 24: \\ 

\par 
-- A similar observation is recorded, with SOL leading once again - 39 trades with a notional value of above \$50,000. 
\par -- By narrowing down on those with trades above \$100,000, the following markets are observed with the maximum recorded in the SOL-USD market on May 23. In fact, that day presented 4 successive trades of abnormally large notional values, with 1 trade '2023-05-23T16:37:47.530Z'  recording over \$500,000.

\begin{table}[h]
\centering
\caption{Number of trades above the amounts - \$50k and \$100k}
\begin{tabular}{|c|c|}
\hline
Trades above \$50,000 & Trades above \$100,000\\
\hline
\includegraphics[width=0.5\linewidth]{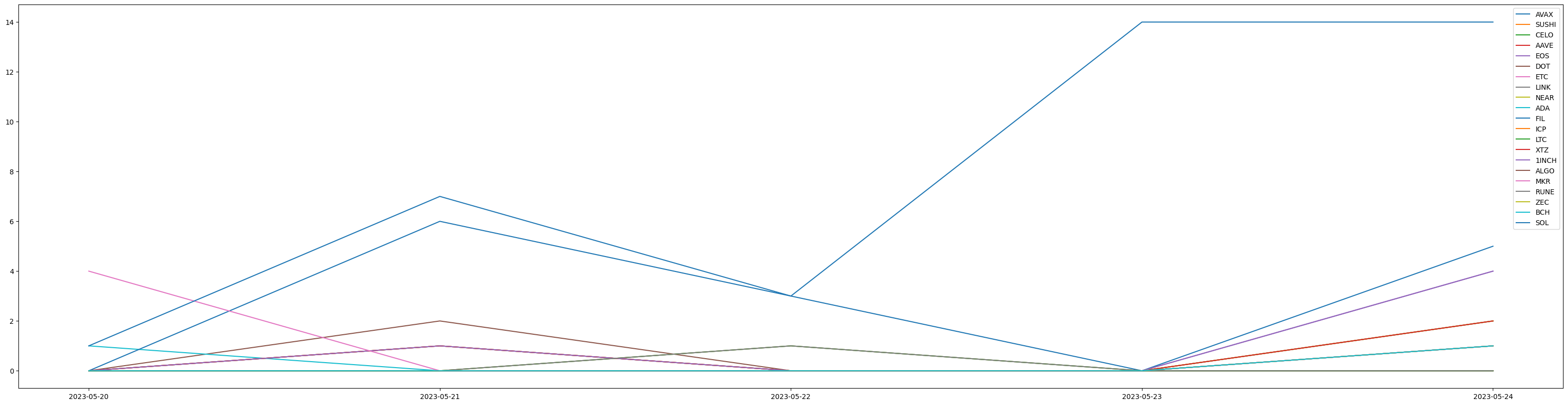} & \includegraphics[width=0.5\linewidth]{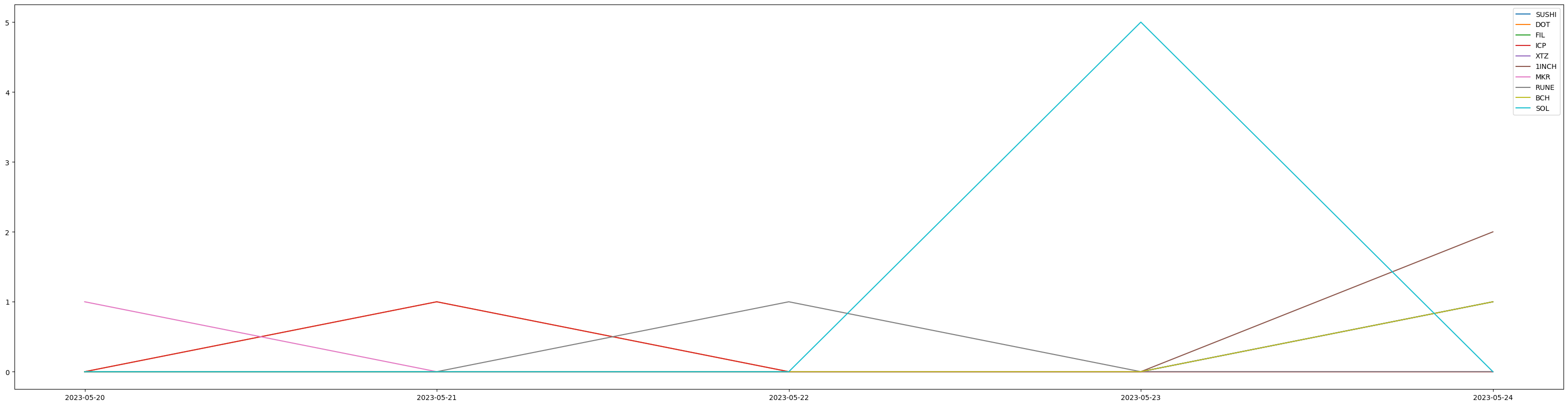} \\

\hline
\end{tabular}
\end{table}

\begin{table}[!ht]
    \centering
    \caption{Trades in SOL USD on May 23}
    \begin{tabular}{|l|l|l|l|l|l|l|}

    \hline
        side & size & price & createdAt & liquidation & day & notional \\ \hline
        BUY & 14087.6 & 20.054 & 2023-05-23T16:38:45.563Z & False & 2023-05-23 & 282512.7304 \\ \hline
        BUY & 10486.5 & 20.054 & 2023-05-23T16:38:37.680Z & False & 2023-05-23 & 210296.27099999998 \\ \hline
        BUY & 29260.0 & 20.059 & 2023-05-23T16:37:47.530Z & False & 2023-05-23 & 586926.3400000001 \\ \hline
        BUY & 19506.7 & 20.045 & 2023-05-23T16:36:38.840Z & False & 2023-05-23 & 391011.80150000006 \\ \hline
        SELL & 5428.0 & 19.97 & 2023-05-23T03:00:55.158Z & False & 2023-05-23 & 108397.15999999999 \\ \hline
    \end{tabular}
\end{table}

\newpage
\par Overall, there are very few 'large sized' orders, suggesting the lack of 'anomalous' trades which may severely dent liquidity. However, extra caution has to be paid to the SOL market given its historically higher volumes recorded on dYdX and larger trade sizes executed.

\subsection{Event Study: High Market Impact Events}

In the aforementioned highlighted high market impact events, an event study was conducted to check the liquidity depths against the reconstructed order book at the same interval. By merging both order book and trades data into a common dataframe at a minute interval, this will feature the order book behavior, contextualized against the trades which occurred.

\subsubsection{Orderbook Resiliency}

\par Based on the derived maxSpreads from the earlier section, these are then visualised and plotted against the high market impact events, to observe \(liquidity_{pre}, liquidity_{post}\)  of market shocks. The duration will be critical to understand how LPs respond to these changes in liquidity and estimate the liquidity replenishment process for the market's self-correcting ability. A threshold of 75\% is set where the estimated time taken for the initial effect of the liquidity shock to dissipate, and liquidity to return to 75\% of the average level is calculated. \\

\par-- Consistent with literature, liquidity generally recovered much faster at wider spreads during these times of volatility, highlighting that LPs and traders are positioning further away from mid price. \\

\par -- For majors, liquidity recovered within 1 and 2 minutes for ETH and BTC respectively for 20bps and below, suggesting that these books are robust and mature enough to attract liquidity in a short period of time.\\

\par -- Markets (LINK, ATOM, ALGO, MATIC, ETC, AVAX, LTC, NEAR) show a consistent quick rate of recovery at 20bps and 30bps in under 3 minutes. \\

\par -- Meanwhile, the remaining markets exhibit some variation in their recovery times. For instance, SUSHI, MKR and ICP takes over 10 minutes to recover at 20bps and below but at 30bps, these assets enjoy a less than 5 minute recovery. On the contrary, ENJ and ZRX see a slower rate of recovery, between 7 and 15 minutes, suggesting that the rate of liquidity uptake far exceeds the liquidity provision under these circumstances. This is similar to the findings of Anand and Venkataraman (2012) where liquidity provision in smaller stocks can be 'sparse and opportunistic'. These results also point to the suggestion to establish DMMs with a strong presence in these less actively traded markets to improve resiliency.  \\

\par * The relevant bar charts can be found in the Appendix. 

\subsubsection{Reconstructed Depths vs maxSpread}

The reconstructed depths were compared with each maxSpread (in intervals between 5bps to 50bps). Should the reconstructed depth exceed the order book depth then, this would suggest that the depth at that particular maxSpread will be insufficient and hence, a higher maxSpread will be checked against to ensure sufficient liquidity buffer. This iterative process is repeated for all markets.

\par \begin{table}[h]
\centering
\caption{Historical Orderbook vs Trades (Reconstructed) for FOMC Meeting Minutes Release (t-1, t+1)}
\begin{tabular}{|c|c|}
\hline
ETH-USD & LTC-USD\\
\hline
\includegraphics[width=0.4\linewidth]{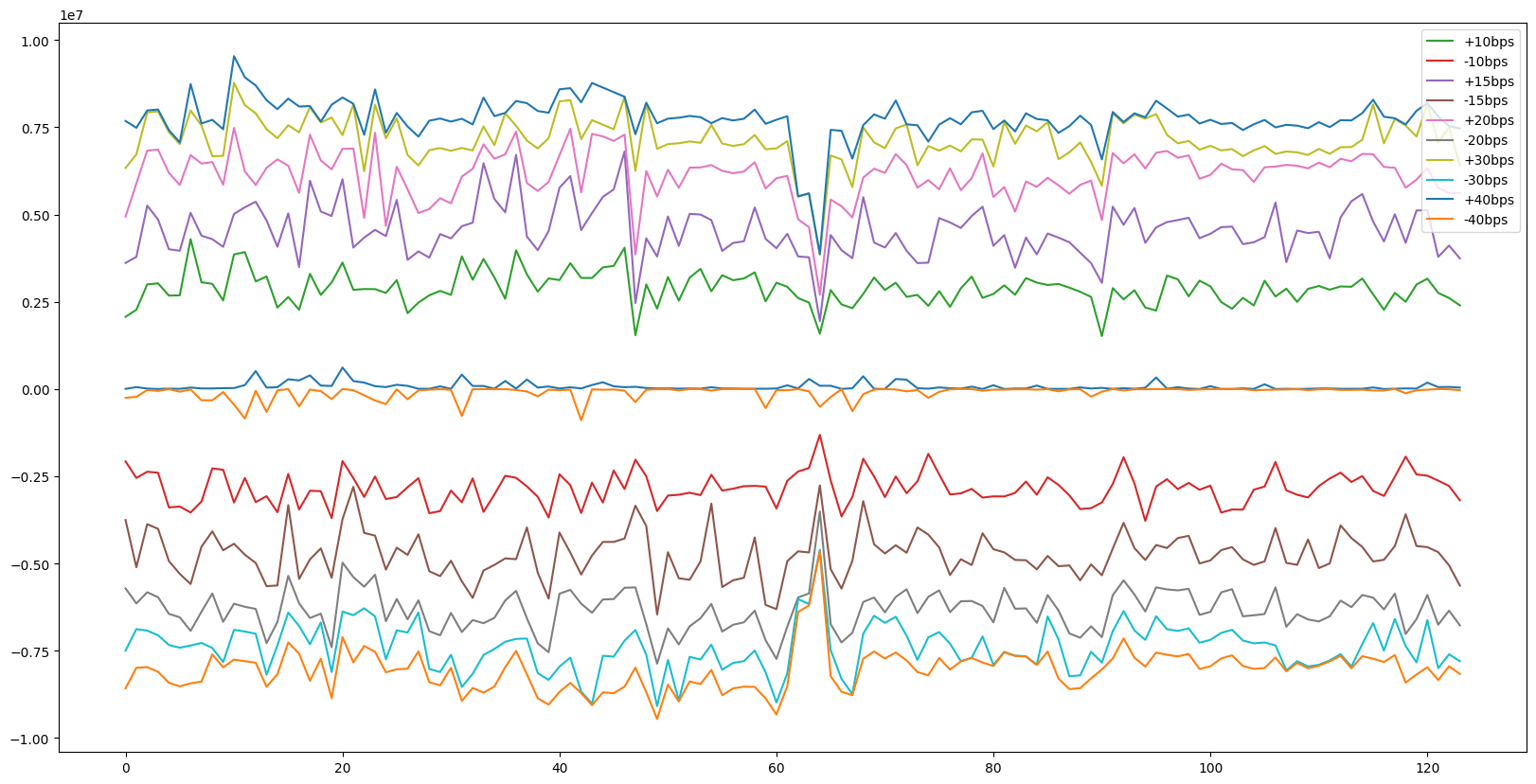} & \includegraphics[width=0.4\linewidth]{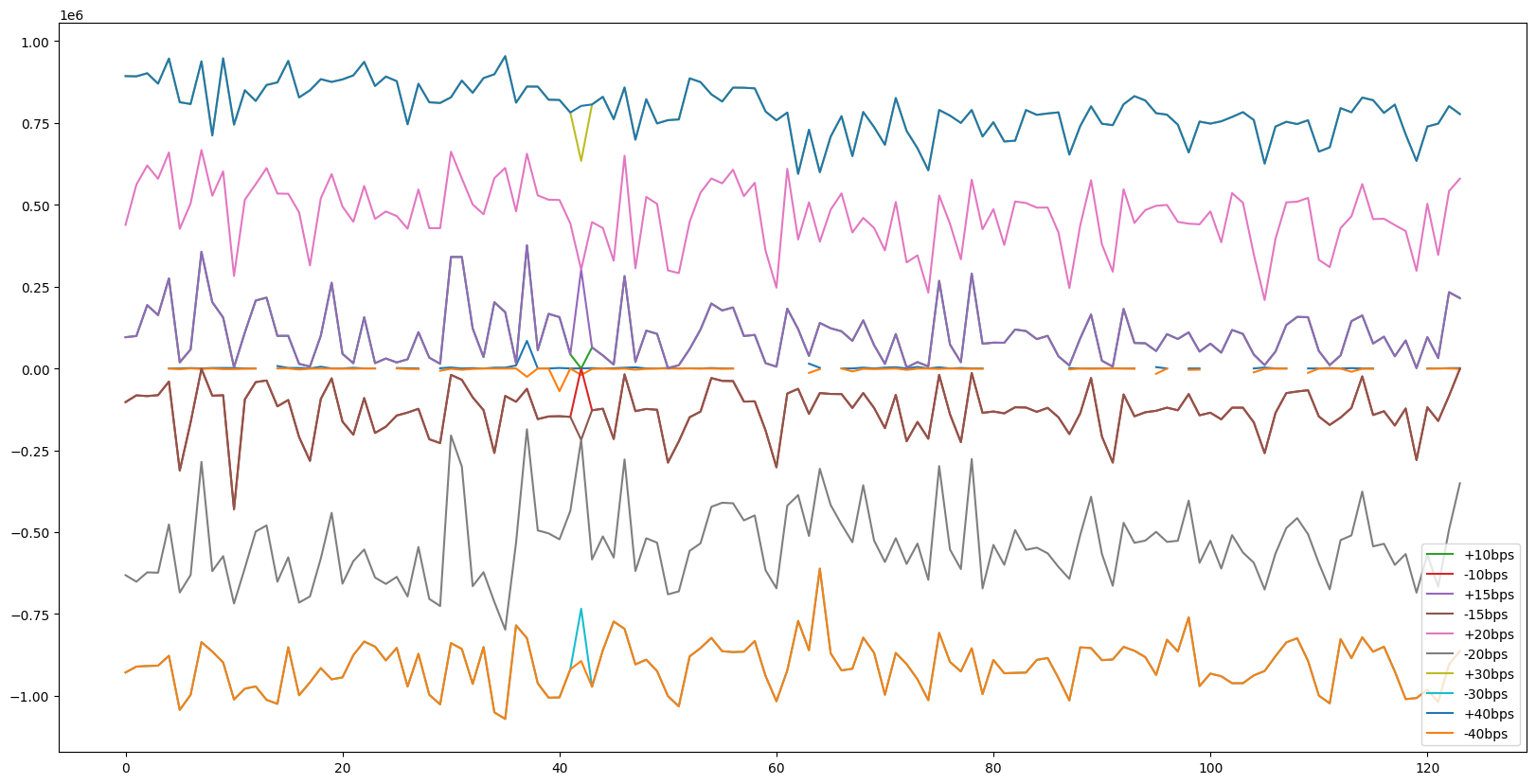} \\
\hline
\end{tabular}
\end{table}

\section{Discussion}

Markets exhibited a distinct segmentation in behavior based on liquidity provision and volume. By establishing the relevant lower bounds (based on tick sizes) and upper bounds (based on reconstructed orderbooks), it can be observed that the exchange had a sufficient amoun tof liquidity in the orderbooks to handle trade flows. This is further supported by the trade data where few huge sized orders are recorded. In particular, markets on dYdX can be further classified into the following categories based on theses characteristics - tick constraints, liquidity and price impacts: 

\subsection{Market Observations}
\subsubsection{Tick Constrained Markets}

Based on Condition 1, markets identified to be tick constrained include UMA and ZRX where a 1 tick difference already incurs more than 40bps. These make it more difficult to trade for LPs given the higher spread costs incurred. For instance, at a 1 tick difference, UMA records 42.8bps for a quote, beyond the maxSpread defined.  As a result, due to the lack of order flows and economic costs, these can remain status quo in the 40bps bracket to compensate the risk borne by LPs by possibly quoting at BBO. 

\subsubsection{Available liquidity}

By observing the historical orderbook snapshots, markets such as COMP, ADA and ENJ recorded similar depth levels at 30bps and 40bps on majority of the occasions, thus remaining in the upper limits of 40bps. \\

On the other hand, more mature altcoin markets with higher volumes and interest can have tighter maxSpreads. It is noteworthy to recognise that majority, if not all of these orders can be absorbed within 20bps of each side on normal days. However, some form of variation exists in their demands and order book response especially from the high market impact events. Markets such as DOT, FIL, CELO, RUNE, DOGE, 1INCH and MKR exhibit rather thinly-traded books especially during these volatile periods, with a distinct difference in liquidity at each depth level. For instance, the ICP market had a significant number of instances where the depths were less than \$50,000 at a tighter maxSpread (30bps). Therefore, these are likely to result in greater price impacts which can be seen from the longer duration in the liquidity recovery process. In the meanwhile, order books of LINK, AVAX, ATOM, ETC, UNI AND LTC were able to sustain sudden influxes in volumes. These markets have also shown strong self correcting abilities to respond to shocks and their liquidity can easily cushion against these events. However, for SOL and MATIC, an evident dent is observed and 30bps may be generally preferred, especially since these are the top 5 most traded markets on dYdX, by historical volume.  Therefore, sufficient liquidity buffer is necessary to dampen potential market shocks.
 
 \subsubsection{Majors : BTC, ETH}
10 bps would be ideal given the maturity of BTC and ETH markets. Historical data has also revealed the consistent tight spreads around the mid price, with deep books on each side. However, given the relatively higher trading volumes on dYdX, 15bps is suggested to provide more buffer. Nonetheless, I will contend in the next section that this part of the formula can be removed.

\newpage
\subsection{Suggested maxSpread parameters}

\begin{table}[h]
\centering
\caption{Suggested maxSpread Parameters}
\begin{tabularx}{\linewidth}{|X|X|X|X|}
\hline
\textbf{Number of Markets} &\textbf{Markets} & \textbf{Original maxSpread parameters}& \textbf{Revised maxSpread parameters} \\
\hline
2 & BTC, ETH &20bps & 15bps  (suggested to be removed in Section 7.1) \\
\hline
6& ETC, LTC, AVAX, LINK, ATOM, UNI& 40bps & 20bps\\
\hline
15 & MATIC, SOL, NEAR, CRV, SNX, YFI, XTZ, XMR, SUSHI, ALGO, XLM, AAVE, BCH, DOGE, TRX& 40bps & 30bps\\
\hline
14& MKR, ICP, COMP, ENJ, EOS, 1INCH, ZRX, UMA, ZEC, FIL, CELO, DOT, RUNE, ADA & 40bps & 40bps\\
\hline
\end{tabularx}
\end{table}

Note: Markets such as UMA remain in the 40bps bracket since 1 tick away is essentially over 40bps. \\

Note:  MATIC and SOL were shifted to the 30bps bracket given its demand on dYdX and the general lack of liquidity depth at 20bps during key events. Nonetheless, there is more than sufficient liquidity to handle daily trades. \\

Assets in in each basket should be re-evaluated monthly, due to differing liquidity conditions. The tick sizes should also be re-calibrated to enable better operating spreads. Nonetheless, the exchange and community should consider implementing a rebates scheme to encourage tighter spreads, and the liquidity rewards to supplement these rebates by incentivising deeper books in the long term. 

\newpage
\section{Further Explorations for Alternative LP Reward Mechanisms}

\subsection{Suggestion 1: Simplify the LP reward formula for mature markets to a volume parameter}

The consistent volume and liquidity depths that these 2 markets attract are testament to the maturity of BTC and ETH in receiving healthy flows. This follows the previous proposal where the weights for makerVolumes were updated to 0.85, and there hasn't been an overt compromise in depth levels on the exchange. A sample from the US Initial Jobless Claims revealed that there was sufficient liquidity available in both ETH and BTC markets. In this case, an average of \$3.5m in BTC and \$4.3m in ETH of liquidity stands within 15bps of the midPrice and these were quick to recover to average levels within 2 minutes. \\

\par \begin{table}[h]
\centering
\caption{Historical Orderbook vs Trades (Reconstructed) for US Initial Jobless Claims}
\begin{tabular}{|c|c|}
\hline ETH-USD & BTC-USD\\
\includegraphics[width=0.5\linewidth]{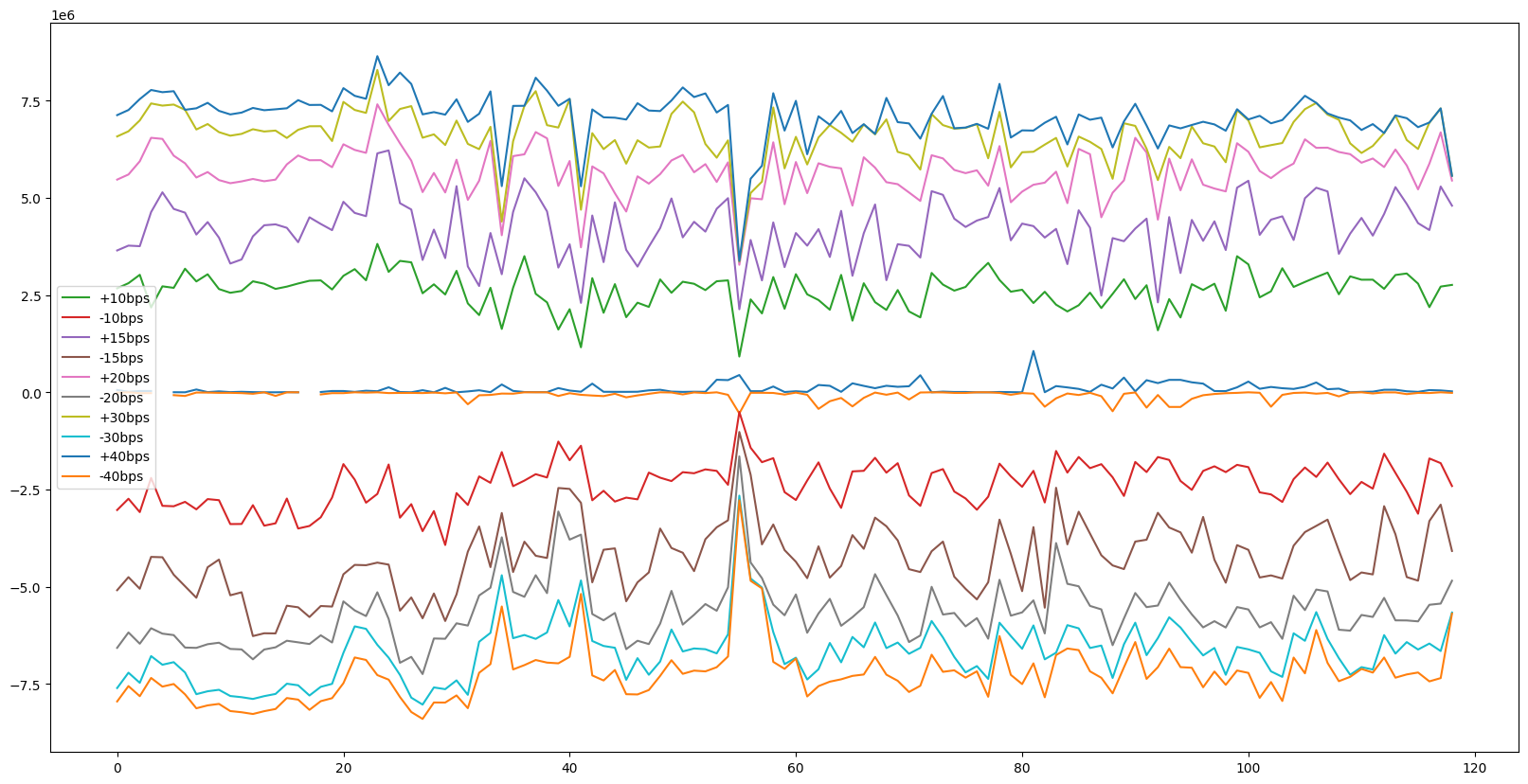} & \includegraphics[width=0.5\linewidth]{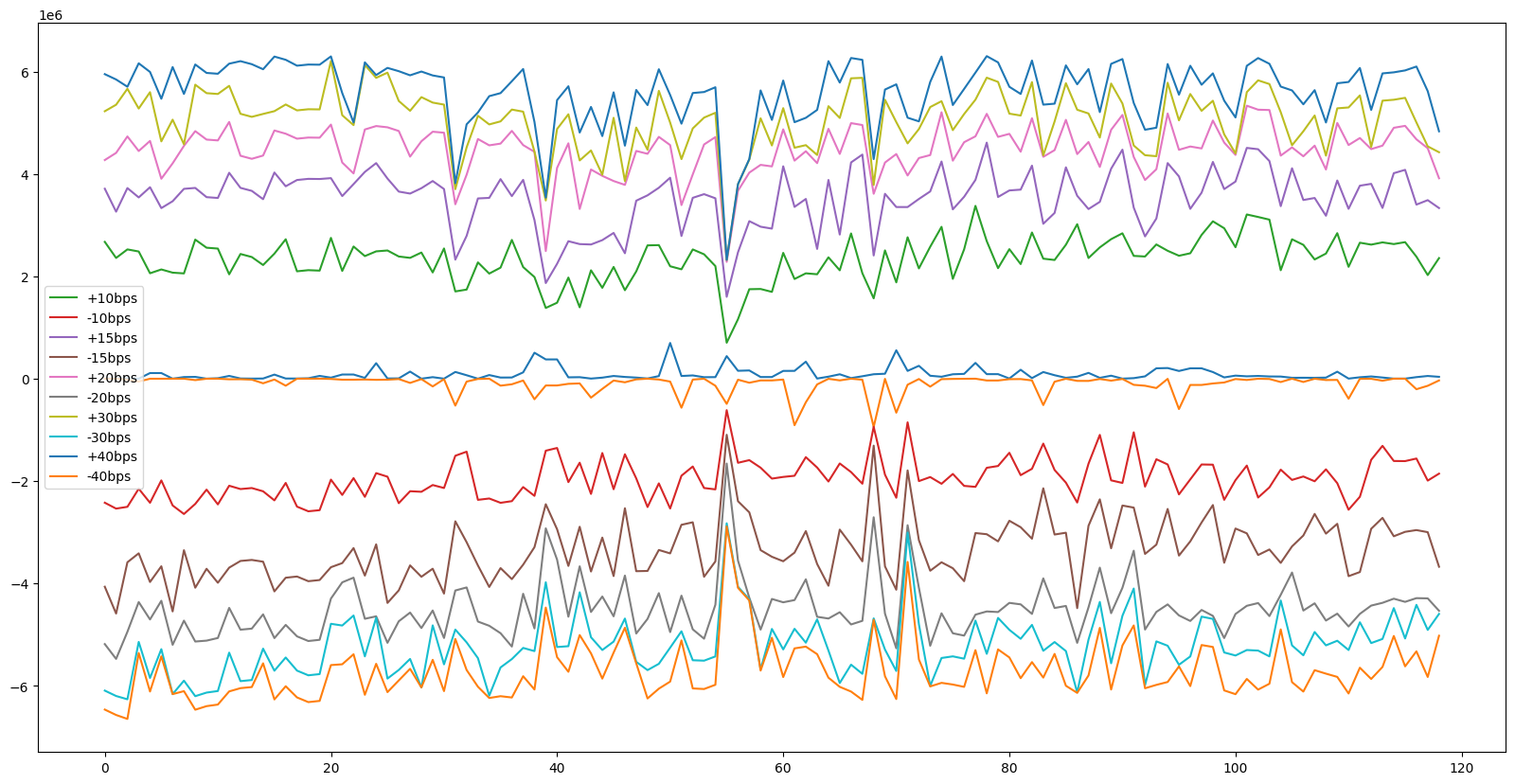} \\
\hline

\end{tabular}
\end{table}

\vspace{10pt}
\par Another observation is that LPs churning far lower volumes still continue to receive a higher amount of rewards. Despite producing 15\% of makerVolume, the account (0xa615) only received 4.67\% of the rewards, much lesser than the accounts above which $produced \leq 10\%$, (mostly hovering around 5 to 7 \% for BTC or as low as 2 \% for ETH markets).\\

Epoch 22: May 9 (BTC Rewards)\\
\begin{tabular}{cccc}
\textbf{Account } & \textbf{Reward Share} & \textbf{Maker volume} & \textbf{Uptime (\%)} \\
0xd332 & 38.2 & 25.67 & 95.4 \\
0xdade & 14.22 & 9.85 & 99.3 \\
0xc698 & 12.17 & 7.28 & 98.4 \\
0xfe85 & 6.78 & 6.5 & 91.9 \\
0xddc9 & 6.16 & 5.33 & 94 \\
0x66e4 & 4.98 & 3.28 & 97.6 \\
0xa615 & 4.67 & 15.39 & 75.5 \\
\end{tabular}\\
\\

Epoch 22: May 9 (ETH Rewards)\\
\begin{tabular}{cccc}
\textbf{Account } & \textbf{Reward Share} & \textbf{Maker volume} & \textbf{Uptime (\%)} \\
0xd332 & 36.18 & 21.47 & 97.4 \\
0xc698 & 16.86 & 11.23 & 96.7 \\
0xdade & 15.33 & 11.07 & 99\\
0xddc9 & 6.38 & 5.48 & 94.7 \\
0x66e4 & 4.66 & 2.76 & 98.8 \\
0x2002 & 4.14 & 2.89 & 96.5 \\
0xa615 & 4.12 & 14.85 & 74.8 \\
\end{tabular}\\
\\

\par While it can be argued that the other accounts had a much higher uptime (more orders within the maxSpread parameter), they have disproportionately lower volumes which suggest that these orders are not near the BBO. In particular, a primary point of contention is that there should no longer be the need for the exchange to still pay for passive liquidity in these mature markets given the demand across exchanges. LPs have naturally provided liquidity within the maxSpreads and (even beyond) to compete for order flow as seen from how quick liquidity recovery occurs around the BBO during high market impact events. Furthermore, by incentivising volumes linearly, this would mark the first step to aligning the scheme with CEXs, before gradually introducing it to other markets.\\

As such, the formula for BTC/ETH markets can be adjusted to:\\

$Q_{x} = \text{makerVolume},  x = BTC, ETH $\\

\emph{Further Implementations:} 
Check for wash trading between LPs to ensure that they are not trading between each other to churn maker volumes. Given the fewer number of LPs relative to takers, this would be feasible implementation by the team in designing the algorithm to catch these accounts. This includes graph techniques to identify strongly connected components in the trade graph and matching of position changes across accounts in the interval. The suggested pseudocode and explanation can be found \href{https://arxiv.org/pdf/2102.07001.pdf}{here}. Alternatively, other suggestions can be found in this \href{https://papers.ssrn.com/sol3/papers.cfm?abstract_id=3530220}{paper} which highlighted 3 main suggestions - Using Benford's law to examine the distribution of the first significant digits in the trade data set, Clustering of transaction sizes at round numbers and check if the observed trade size distributions have fat tails based on the power law distribution.

\subsection{Suggestion 2: Implementing the LP Rebate Scheme}
Wintermute previously introduced a  successful \href{https://snapshot.org/#/dydxgov.eth/proposal/0x8d7b06b21e50756479a4738924039adf74705d66d8c92554cff74dbab52a9a73}{snapshot} - 'Maker Maker Rebate Programme to reduce the reliance on LP rewards and incentivise liquidity in a natural way'. This would mean that LPs' rewards are now 'directly proportional to their maker volume', a similar measure by many exchanges. Therefore, one can easily quantify the rewards that can be possibly earned by LPs and even conserve the allocated DYDX rewards for future use (such as boostrapping new markets). Furthermore, LPs are given certainty since rebates are denominated in USDC, reducing delta exposure to DYDX prices. \\

However, one has to consider if the rebate scheme is sufficient to overhaul the LP scheme. Based on Epoch 22, there were a total of 74 LPs with a volume $\geq$ 0.25\% of the exchange's volume. Using the suggested fee schedule, 19 LPs will be eligible for Tier 5 as they collectively provide for 71.4\% of exchange's volume. At a 0.0100\% rebate, this equates to \$2,164,061 in USDC already given to this group of LPs. In comparison, the LP programme emits  \$2,301,370 (1,150,685 DYDX tokens at \$2 each) per epoch. \\

\begin{figure}[htbp]
    \centering
     \caption{Epoch 22 Maker Volume Distribution}

    \includegraphics[width=0.8\textwidth]{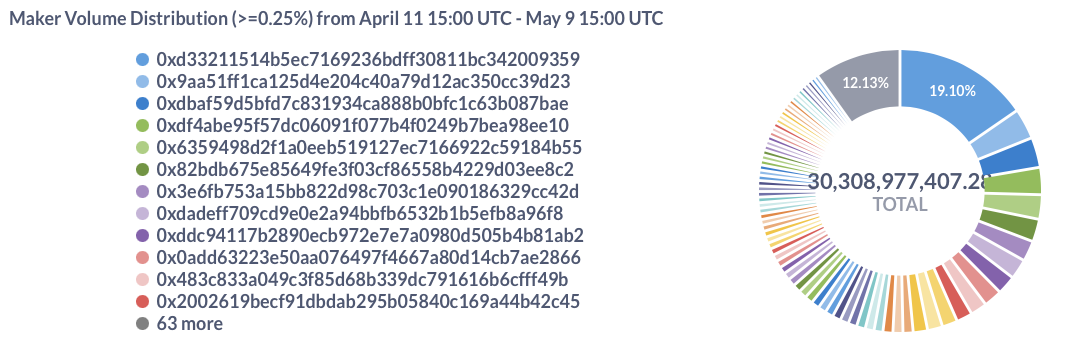}
    \end{figure}

\par In this case, it is important to recognise that majority of the volume is driven by the majors\footnote{Majors = BTC, ETH}(\~80\%). As such, a pure rebate scheme for the altcoin markets will likely prove ineffective since these rebates will be significantly lower than the present LP incentives. In Epoch 22, altcoin volumes recorded ~\$6,276,767,740.46. Assuming all LPs who provided in these markets receive the maximum 0.0100\% rebate, this is equivalent  to \$627,676.77 (current LP allocation for altcoins = \$1,841,096.0 at \$2 each). This can be observed from the following simulation:\\

\par a. Altcoin fee revenue is approximately \$1.13M, Altcoin Volume is estimated to be \$6.3B. \\
\par b. This equates to 0.0179\% earned by the exchange through the current maker and taker fees. Therefore, an upper bound of rebates is set at this rate to ensure the exchange remains profitable. Likewise, the lower bound of rebates will be at the highest tier in Wintermute's proposed fee schedule of 0.0100\%.  \\
\par c. Plotting out the volumes, rebates and present LP rewards scheme, it can be seen that the rebates are relatively lower in value, unless volumes scale above \$11B. \\

\begin{figure}[ht]
  \centering
  \caption{Rebates vs Rewards (For Altcoins)}
  \includegraphics[width=0.8\linewidth]{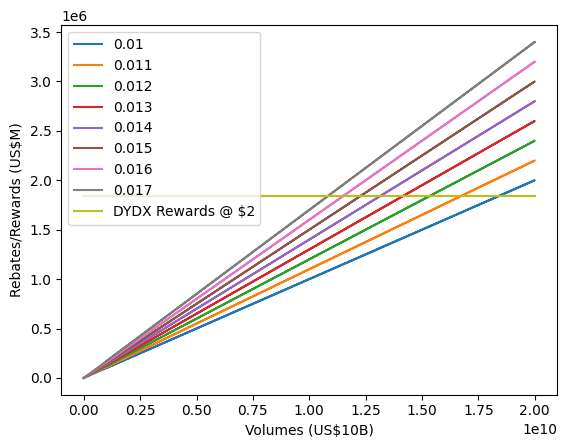}
  
\end{figure}

\newpage

Henceforth, this will require a careful calibration to implement the rebate scheme and adjust the rewards segment moving forward. After all, since this will directly incentivise active liquidity, the changes in depth will have to be closely monitored to ensure deep liquidity still exists.\\

\emph{Consideration: Lower Exchange Revenue but more DYDX Tokens.} Presently, most of the fee revenue is driven by the majors with \$3,472,552.65 while the altcoins rake in \$1,128,679.07\footnote{18 May - 17 June as this data was added in at a later date}. The next 2 suggestions will be focussing on possible rebates schemes but these will result in a contraction in overall revenue, should volumes not increase proportionally. 

\subsubsection{Replace the LP Incentive Programme for BTC and ETH with the Rebates Scheme}

Should the formula be linearized to a volume parameter for the majors, then the rebates scheme can be implemented, re-allocating the 20\% of DYDX rewards. \\

Based on epoch 22, BTC and ETH recorded \$\~25,059,174,253 in volume. \\

1. The volumes can be rounded off to \$25B for calculation. \\
\par 2. Using the suggested fee schedule and the previous epoch data (where 19 LPs are eligible), they provide for an estimated 90\% of volumes for the majors. \\
\begin{table}[htbp]
\centering
\caption{Maker Volume Distribution for BTC and ETH}
\begin{tabular}{cc}
\includegraphics[width=0.4\linewidth]{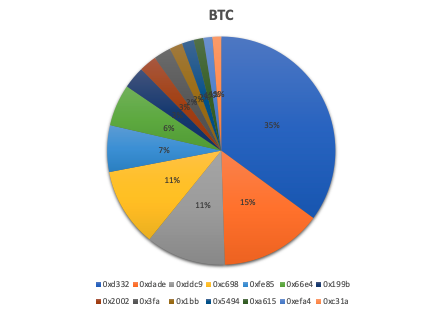} & \includegraphics[width=0.4\linewidth]{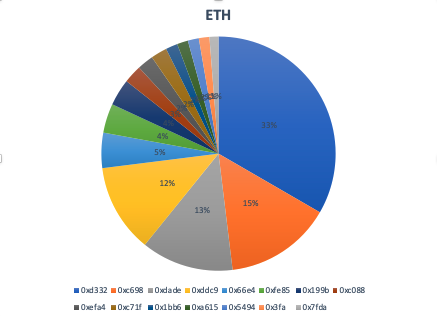} \\
\end{tabular}
\end{table}
\par ** Other LPs with $\leq 1\%$ volumes are not included in these pie charts.\\

\par 3. With the 0.01\% fee rebate, this equates to approximately \$2.25M of USDC rebates (\$25B * 90\% * 0.01\%), significantly higher than the present DYDX allocated to these markets (230,127 DYDX Tokens = \$460,274 at a price of \$2). \\
\par 4. Over time, as more LPs are attracted to provide liquidity, this may level off the playing field in volume distributions, encouraging the formation of a more healthy and mature market.\\

Supported by the previous Section 7.1, the formula can be updated to solely incentivise makerVolumes with a strong self-correcting market and deep liquidity. Since LPs and the community have already signalled the intention that they are agreeable to a rebate scheme, this can be experimented starting from the majors. It not only conserves the DYDX treasury, but LPs are likely to reap better economic benefits and certainties from providing liquidity in these markets. This begets liquidity demand, enabling a virtuous cycle of market activities.

\subsubsection{Enhanced Rebates for long tail / new markets}

Currently, a one-size-fit-all rebate scheme is suggested across the exchange, which may not necessarily incentivise LPs to provide for newly listed and existing thinly traded markets. Inspired by the NASDAQ DLP Programme, an enhanced rebates scheme can be considered. This was used to incentivise LPs to 'support a broader array of ETPs, building the foundation for emerging ETPs' volume..'. A similar programme, on top of the suggested rebate scheme can be implemented to attract LPs and potentially reduce the amount of DYDX rewards. \\

\par This can be seen from the reward coeffecient which displays the number of DYDX earned for a constant level of liquidity (depth and spread) provided. A higher coeffecient will attract LPs to provide liquidity into these markets. However, market such as UMA (4.70), ZRX (3.86), ZEC (3.56), ENJ (2.67) have high coeffecients based on the past 3 months and yet they experience relatively thin orderbooks with deviant spreads. Evidently, LPs are not attracted to provide liquidity in these areas to begin with. \\

\begin{table}[!ht]
    \centering
    \caption {Suggested Rebate Scheme for all markets (By Wintermute)}
    \begin{tabular}{|l|l|l|}
    \hline
        ~ & Maker Volume (\% of 30D Exchange Volume)  & Normal Rebates  \\ \hline
        Tier 1  & $\geq$ 0.1\% & -0.0025\% \\ \hline
        Tier 2  & $\geq$ 0.25\% & -0.0040\% \\ \hline
        Tier 3  & $\geq$ 0.5\% & -0.0050\% \\ \hline
        Tier 4  & $\geq$ 0.75\% & -0.0065\% \\ \hline
        Tier 5 & $\geq$ 1\% & -0.0100\% \\ \hline
    \end{tabular}
\end{table}

\begin{table}[!ht]
    \centering
    \caption{Enhanced Rebates Scheme for Long Tail / New Markets}
    \begin{tabular}{|l|l|l|}
    \hline
        ~ & Maker Volume (\% of 30D Market Volume)  & Enhanced Rebates \\ \hline
        Tier 1  & $\geq$ 5\% & -0.0125\% \\ \hline
        Tier 2  & $\geq$ 10\% & -0.0150\% \\ \hline
             \end{tabular}
\end{table}

Currently, there are very few LPs (within 3 per market) that contribute at least 10\% of the market's volume in these long tail markets and thus, a higher level of rebates may possibly encourage other LPs to move up the tiers and participate in the liquidity provision process. These should be trialled to observe the improvements in orderbook depths while gradually calibrating the incentives scheme. 

\subsection{Suggestion 3: Increase the makerVolume weightage for non-BTC/ETH markets to 0.85}

The present LP rewards formula heavily emphasises on depths and spreads given the 2 components (bid, spread and Qmin) scores, with just 0.65 for makerVolume. Especially with the persistently low volumes and wide spreads, the system can be easily gamed by LPs who provider wider and passive orders which hardly get hit. This drives the scenario where quoting at 40bps away from midPrice for the SOL market still returns a sizeable amount of rewards. \\

A concern is the ratio of volumes for these markets relative to other CEXs. Altcoins vs BTC/ETH on dYdX presents an extremely disoriented picture where there is an overt amount of trading activities in the majors. The lack of volume proves worrying and this can be seen from the relatively wide quotes away from the BBO (earlier snapshot on May 4), causing traders to suffer from worse slippages, deteriorating the experience. In fact, many of these markets offer deeper books relative to the average volumes recorded on dYdX, suggesting that there should be a greater emphasis on makerVolumes to drive active liquidity. \\

In particular, from a LP rewards perspective, similar circumstances are observed where LPs with higher makerVolumes receive a disproportionately lower amount of rewards: \\

\emph{Epoch 22: Top 5 LPs }
\vspace{10pt}\\
\par

\begin{tabularx}{\linewidth}{*{8}{X}}

\textbf{1INCH} & \textbf{Reward} & \textbf{Vol} & \textbf{Uptime} & \textbf{ALGO} & \textbf{Reward} & \textbf{Vol} & \textbf{Uptime} \\
0x66e4 & 42.29 & 32.7 & 99.3 & 0x66e4 & 28.23 & 22.61 & 99.2 \\
0xd332 & 14.25 & 13.69 & 98.2 & 0xc698 & 26.16 & 16.15 & 98.9 \\
0xa615 & 6.45 & 13.56 & 98.1 & 0x57c4 & 12.84 & 17.36 & 99.3 \\
0xc698 & 12.38 & 6.73 & 99.5 & 0xd332 & 10.2 & 8.39 & 96.5 \\
0xddc9 & 10.68 & 8.15 & 98.0 & 0xddc9 & 10.19 & 6.77 & 96.5 \\
\end{tabularx}\\
\vspace{10pt}\\
\par 
\begin{tabularx}{\linewidth}{*{8}{X}}
\textbf{AAVE} & \textbf{Reward} & \textbf{Vol} & \textbf{Uptime} & \textbf{ MATIC} & \textbf{Reward} & \textbf{Vol} & \textbf{Uptime} \\
0xd332 & 41.72 & 36.66 & 98.9 & 0xd332 & 30.16 & 24.29 & 98.7 \\
0x66e4 & 20.88 & 18.59 & 98.1 & 0xa615 & 7.56 & 16.72 & 94.7 \\
0xc698 & 13.85 & 9.04 & 98.7 & 0xc698 & 18.62 & 12.91 & 98.8 \\
0xddc9 & 11.25 & 8.38 & 97.6 & 0x66e4 & 14.18 & 13.91 & 98.4 \\
0xa615 & 4.1 & 7.18 & 96.5 & 0xddc9 & 9.68 & 7.45 & 97.5 \\
0xfe85 & 1.13 & 10.57 & 63.7 & - & - & - & - \\
\end{tabularx}
\vspace{10pt}\\

\par 
\begin{tabularx}{\linewidth}{*{8}{X}}
\textbf{SOL} & \textbf{Reward} & \textbf{Vol} & \textbf{Uptime} & \textbf{ADA} & \textbf{Reward} & \textbf{Vol} & \textbf{Uptime} \\
0xd332 & 48.02 & 41.03 & 98.9 & 0xc698 & 25.98 & 15.79 & 98.7 \\
0xc698 & 16.99 & 9.98 & 98.2  & 0xd332 & 18.95 & 9.06 & 98.4 \\
0xddc9 & 9.59 & 6.98 & 97.9  & 0xa615 & 17.72 & 29.13 & 97.3 \\
0xa615 & 7.67 & 15.88 & 93.2 & 0xfe85 & 8.73 & 4.41 & 97.8 \\
0xefa & 4.66 & 3.87 & 89.1 & 0xbce6 & 6.74 & 17.03 & 90.7 \\
\end{tabularx}

\vspace{10pt}

Despite producing nearly 16\% of makerVolume in SOL, account 0xa615 has only received 7.67\% of rewards which is disproportionately lesser than the above account with over half of the makerVolume. This observation is similarly seen across multiple markets such as 1inch, AAVE, MATIC and ADA (as seen in the tables above). While sufficient liquidity buffer is necessary, makerVolumes should also be incentivised to provide tighter quoting. Some of these altcoin markets are also primarily dominated by the same group of LPs and would have already witnessed a relatively high uptime. Thus, a performance breaker would be focussing on the makerVolume to reward those for providing liquidity in the active ranges. This suggestion has witnessed similar success previously with the BTC/ETH markets where the weightage rose to 0.85 and hence, should be trialled in the altcoin markets. \\

The updated weightage should be ceded for discussion. However, I would recommend this to be only implemented after the maxSpread parameters have been revised to ensure that depths have not been not overly penalized.
 
\subsection{Suggestion 4: Assigning DMMs for markets per epoch}

DMMs are a common implementation in traditional exchanges where they are appointed by the exchange to provide liquidity in certain markets based on a set of criteria. This may include a minimum BBO presence, minimum passive volume share and passive volume ratio. For instance, these DMMs will be required to provide orders for a percentage of their trading time within the BBO while hitting a required share of their passively executed volume which remains in the order book. A similar programme can be implemented for the different markets. \\

The main rationale would be that every LP has a particular focus on certain markets based on the products they are proficient in. Especially in v4, where permissionless market listings become possible, this provides an avenue for LPs who specialize in these new markets to actively provide liquidity. This phenomenon can already be observed on dYdX where the markets are primarily dominated by a group of different LPs which record disproportionately higher makerVolume and rewards (as seen in Section 7.3). Furthermore, the irregular orderbook depths can be observed in the long tail markets during high market impact events.\\

\begin{table}[h]
\centering
\caption{Historical Orderbook Depth Snapshot from FOMC Minutes Meeting Release (at 40bps)}
\begin{tabular}{|c|c|}
\hline
ADA-USD & ENJ-USD \\
\hline
\includegraphics[width=0.4\linewidth]{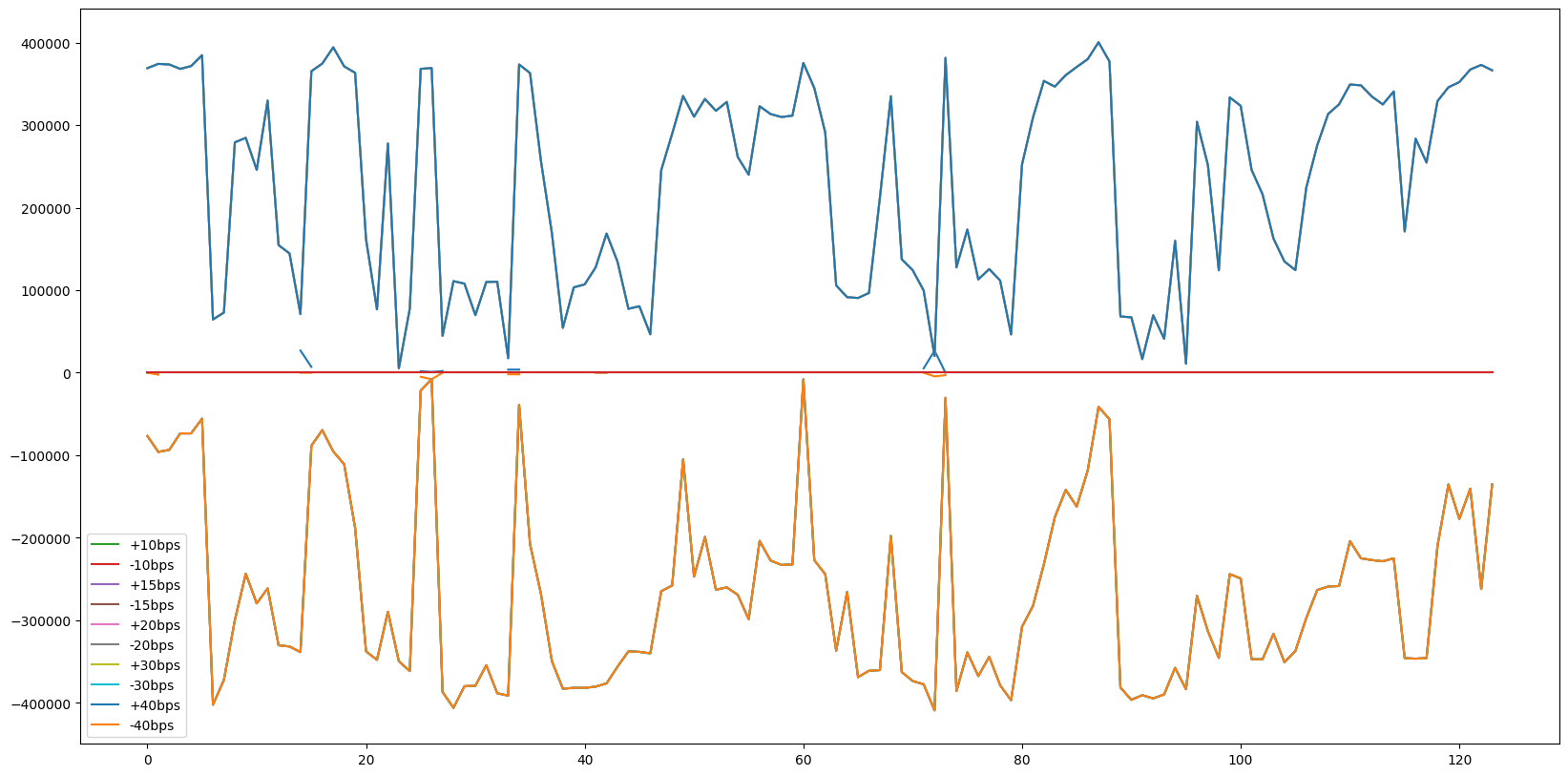} & \includegraphics[width=0.4\linewidth]{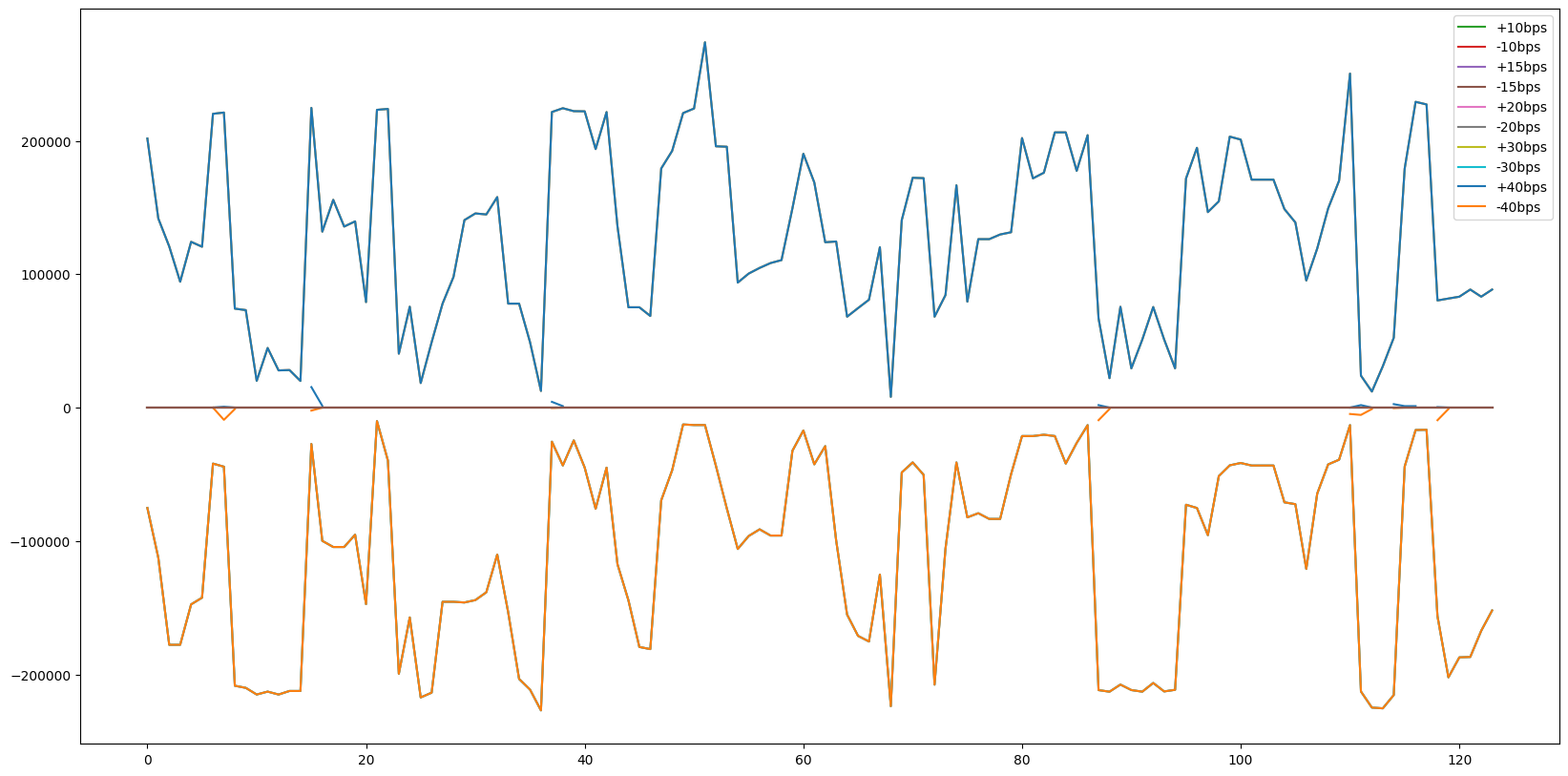} \\

\hline
\end{tabular}
\end{table}

Therefore, to compensate LPs for the risk of trading in these  new / 'higher-risk' markets, a trial can be established in the low volume altcoin markets to model the likely improvement in liquidity. A suggested model can be seen below: \\

\begin{figure}[ht]
  \centering
  \caption{DMM Structure Per Epoch}
  \includegraphics[width=1.1\linewidth]{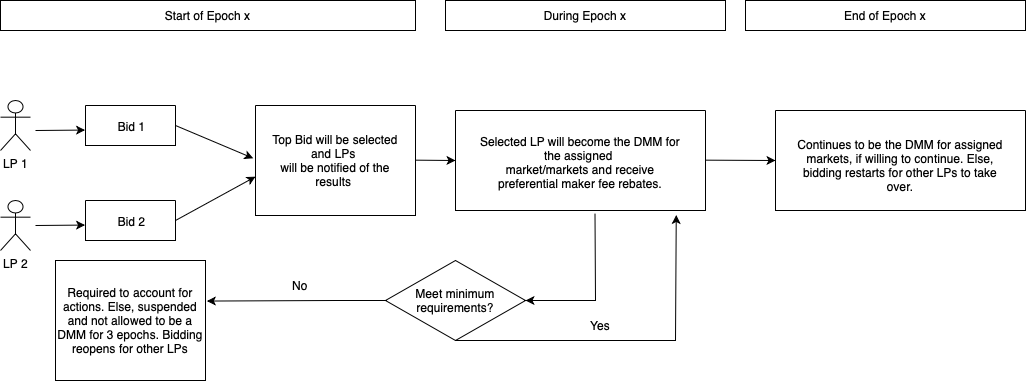}
  
\end{figure}

\newpage
A minimum requirement can be implemented for existing and new markets where the weighted score will be calculated for each LP:\\

\begin{table}[!ht]
    \centering
    \caption{Suggested DMM Bidding Requirements}
    \begin{tabular}{|l|l|l|l|}
    \hline
        ~ & Conditions & Existing Market & New Market \\ \hline
        1 & Number of Epochs the LP has been providing liquidity  & Y & Y \\ \hline
        2 &  $\geq$  2\% of total volume traded  & Y & Y \\ \hline
        3 & Top 3 by makerVolume share OR  $\geq$  20\% of makerVolume  & Y & Y \\ \hline
        4 &  $\geq$  95\% Uptime  & Y & Y \\ \hline
        5 & Top 3 by Per Minute Depth-Spread Score & Y & Y \\ \hline
        6 & Amount of liquidity provision intended (Commitment)  & x & Y \\ \hline
        7 & maxSpread relative to midPrice (Max of 40bps)  & x & Y \\ \hline
        8 & Max Price Deviation (Volatile Events) & 10\% & 10\% \\ \hline
    \end{tabular}
\end{table}

** Condition 3 : Whichever condition entails a lower makerVolume. For instance, certain markets may be dominated by a LP providing over 50\% of makerVolume. Hence, to encourage more LPs to participate, the parameter is widened. \\

** Since these are new markets with uncertain volumes, DMMs are expected to provide a minimal amount of liquidity into the market and assist in the price discovery process. maxSpreads should be ideally as tight as possible to facilitate flows. Therefore, these are additional requirements above the usual metrics. \\
\\
** These numbers are based on the scores observed on the long tail markets.\\

\[Score_{LP} = \sum_{i=1}^{n} \frac{x_{i}}{total_{i}}\]

\begin{table}[!ht]
    \centering
    \caption{Sample Calculation for Average Weighted Scores}
    \begin{tabular}{|l|l|l|l|}
    \hline
        ~ & Number of Epochs  & \% of Total Volume Traded & .. \\ \hline
        LP 1  & 10 & 30 & .. \\ \hline
        LP 2  & 6 & 10 & .. \\ \hline
        Total Number & 16 & 40\% & ~ \\ \hline
    \end{tabular}
\end{table}

Therefore, LP 1's score = $(10 \div 16) + (30 \div 10) +$ .... \\

Notations: M = Minimum Assets Requirement on dYdX, A = Penalty factor, R = Reward multiplier 
\subsubsection{Start of Epoch}

Each LP can submit their intended functioning parameters based on the aforementioned metrics, and indicate the markets which they intend to apply for while having this minimum amount on the platform as part of this process. A simplified formula is proposed where $ M =  Liquidity_{Epoch} * 0.02\% $. This is akin to the net liquid assets requirements for DMMs in traditional exchanges. It is critical that the DMM unit maintains sufficient liquidity, to ensure an orderly market in these assigned markets especially during market stress. \\
\\
eg. 
** If the market records a daily average of \$5M, M required = \$5M * 28 * 0.02\% = \$28,000 USDC. \\

With reference to the metrics, should these be existing markets, liquidity will refer to the rolling average for the past 3 epochs of historical volume. For new markets, liquidity will be the amount committed by the LPs based on their bids. Using the equally weighted formula, the LP with the highest score will be the DMM for that market and will be obliged to provide liquidity at their submitted parameters.

\subsubsection{During Epoch}

Weekly reviews will be conducted (e.g. by the  \href{https://dydx.forum/t/crafting-dydxs-decentralised-future-essential-subdaos-for-consideration/215/10}{LP/Validator subDAO} ) to ensure that the DMM meets the minimum obligations. This can be based on the existing LP rewards dashboard visualized on datahog.\\

\emph{Failure to Meet Obligation} - Should the DMM fail to meet their obligations for 1 day, the DMM will be required to account for the discrepancies on the forums. Otherwise, the DMM will be suspended and removed from the programme. It will then not be allowed to participate as a DMM in the subsequent 3 epochs. The cycle repeats again with LPs bidding for the position to replace the DMM. \\

\emph{Successfully Meets Obligation} - During this period, the DMM will be eligible for a different maker fee structure that confers preferential treatment to this group of LPs. In particular, the markets on the exchange can be classified similarly into 'More Active Markets' and 'Less Active Markets' based on their historical volumes and spreads.   \\

In determining the structure, the following equation was used, where the fees and volumes are summed up for all markets in each tier, and assuming 50\% represents the makerVolume share of the DMM: \\
\begin{equation}
  Fee Revenue  -  (Volumes  \times Rebates \times 50\%)
\end{equation}
This would represent the approximated margins earned by DYDX. The shortfall will be compensated with these suggested rewards at a price of \$2, acting as a 'stipend'.
\newpage
\begin{table}[!ht]
    \centering
    \caption{DMM Rebates and Rewards Structure}
    \begin{tabular}{|l|l|l|l|}
    \hline
        ~ & Markets (excl BTC/ETH) & Fees & DYDX \\ \hline
        More Active Markets & SOL, MATIC, LTC, AVAX, ADA, DOGE, ATOM & -0.0125\% & 7500 \\ \hline
        Less Active Markets & XLM, COMP, CELO, ENJ, ZRX, ZEC, RUNE, UMA  & -0.0150\% & 2500 \\ \hline
        Active Markets & Remaining Markets & -0.0150\% & 5000 \\ \hline
        ~ & Total Emissions & ~ & 172500 \\ \hline
    \end{tabular}
\end{table}

** 'More Active Markets' tend to be dominated by the top LPs and will likely be eligible for the 0.0100\% rebates proposed previously and incur 0 maker fees. Hence, raising this to 0.0125\%. The exchange still continues to earn fee revenue given that the lowest taker fee is at 0.0200\%. 

\subsubsection{End Epoch}

Should the chosen LP be willing to continue as the DMM for the assigned market/markets and have successfully met its obligations for the epoch, there will be no need to restart the bidding period. However, if it chooses to discontinue, then other LPs will be invited to bid and take over as the DMM.\\

Ultimately, this opens up a new mechanism for specific LPs to provide liquidity to their preferred products without overtly spreading across multiple markets. Especially for new markets, this aids in the bootstrapping of liquidity by encouraging LPs to aggressively provide better quotes and create a healthy orderbook flow. As a result, improving the market quality given the increased competition through a pragmatic way of enjoying better maker rebates. \\

\emph{Further Implementations:} Create a group of DMMs based on market segmentation (of volume). Since certain markets are primarily dominated by a group of LPs, markets can be grouped together to reduce the complexity of this structure. This will involve the exchange identifying commonalities - eg.  SOL, MATIC and AVAX may share a group of 2 to 3 DMMs subject to the aforementioned criteria. 
 
\subsection{Suggestion 5: Sunset or reduce rewards for certain existing v3 markets}

Another consideration would be to study the profit margins for each market: Fee Revenue - DYDX Rewards allocated. Based on historical observations when benchmarked against other CEXs, we can observe the relatively lower volumes across the space and the minute fee revenues generated from activity in these markets.  Furthermore, while these rewards are distributed based on the number of Qminsamples, it still effectively categorises extremely long tail markets such as UMA into the same basket with the SOL market. A look at the historical volumes (excluding BTC/ETH) highlights the disparity across markets, where SOL reached as high as 12\% of total volume. As such, rather than continuously allocating rewards to these 'low profit' markets, the exchange can better utilize its rewards to drive volume in more active markets. \\

Based on the dYdX API, the fees generated from each market can be seen below (Statistics can be found in Appendix 10.8)\footnote{18 May - 17 June as this data was added in at a later date}: 

\begin{table}[!ht]
\centering
\caption{Fees Generated (30 Days)}
\begin{tabular}{|c|c|}
\hline
All markets on dYdX& Altcoin Markets on dYdX \\
\hline
\includegraphics[width=0.4\linewidth]{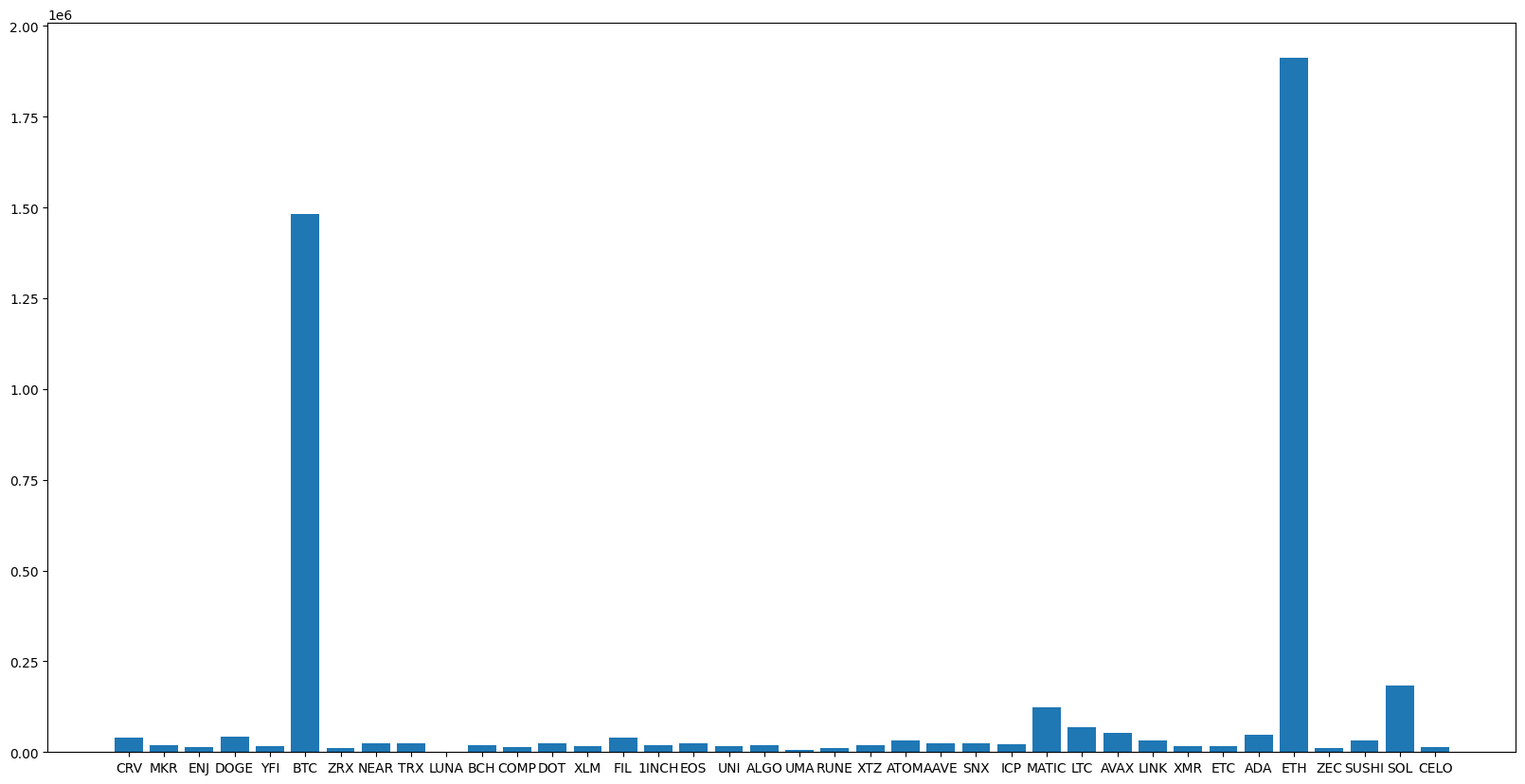} & \includegraphics[width=0.4\linewidth]{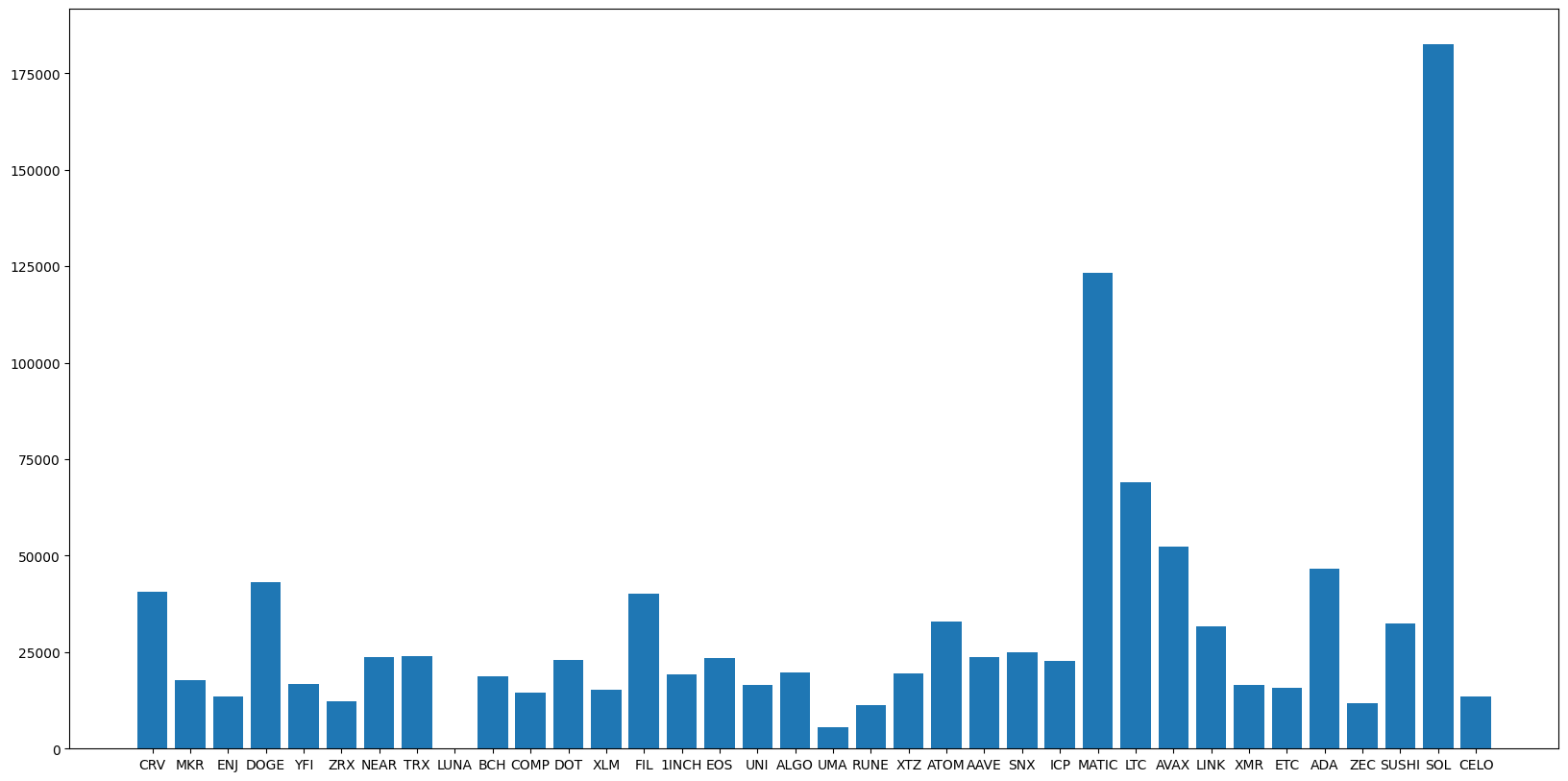} \\

\hline
\end{tabular}
\end{table}

Altcoin markets have brought in a fee revenue of \$1.13M over the 30 days. (This accounts for total fees paid by users - including LPs and takers). In comparison, the 80\% of LP rewards allocated, (at price of \$2), stands at \$1.84M. In deciding the allocations for each market, we can consider the following: \\

\par 1. Historical fee revenue generated from the market should be considered. Should it produce relatively low fees, then the market should receive a lower amount of rewards. \\
\par 2. Markets should only be incentivised if they see continuous demand on the exchange and across other CEXs. A comparison in historical volumes can be done with Binance, Coinbase and Okx and historically low volume market can be considered for delisting.\\

An initial suggestion would be to classify the following markets into these tiers based on the fee revenue and volume. \\

\begin{table}[!ht]
\caption{Suggested Tiers}
    \centering
    \begin{tabular}{|l|l|}
    \hline
        ~ & Markets \\ \hline
        Tier 1 & SOL, MATIC, LTC, AVAX, ADA, DOGE, ATOM \\ \hline
        Tier 2 & CRV, FIL, UNI, LINK, SNX, AAVE, TRX, NEAR, EOS, DOT \\ \hline
        Tier 3 & ICP, ALGO, 1INCH, XTZ, BCH, MKR, XMR, YFI, SUSHI, ETC \\ \hline
        Tier 4 & XLM, COMP, CELO, ENJ, ZRX, ZEC, RUNE, UMA  \\ \hline
    \end{tabular}
\end{table}

\begin{table}[!ht]
\caption{Suggested Allocation }
    \centering
    \begin{tabular}{|l|l|l|l|}
    \hline
        ~ & Fee Revenue & Allocation (if 80\%) & Allocation (if 100\%) \\ \hline
        Tier 1 & 555494.79 & 40\% & 50\% \\ \hline
        Tier 2 & 276592.38 & 20\% & 30\% \\ \hline
        Tier 3 & 199288.70 & 15\% & 15\% \\ \hline
        Tier 4 & 97303.20 & 5\% & 5\% \\ \hline
    \end{tabular}
\end{table}
\newpage

Alternatively, we can consider capping the reward allocations for the Tier 4 markets to 5\% as highlighted above given their historically low volumes and fees generated. \\

However, future studies should be done to provide a granular understanding of the current reward allocation to each of the altcoin markets (should the data be made available in the future). 

\subsection{Other Suggestions:}

\subsubsection{Expand the LP Rewards Formula - Reward Multiplier for Volatile Periods}

\emph{Volatile periods can be defined as periods of time where prices fluctuate, presenting much uncertainty due to asymmetric information from a mixture of informed and uninformed traders.}\\

In markets with high volatility / low volume, investors and LPs will want to place orders far away from trading levels to benefit from violent movements. This results in a fatter tail as the spreads are noticeably wider. LPs can also simply withdraw their liquidity during that short period of time, which does not really affect their uptime score. Hence, LPs with a high uptime during the volatile periods should be adequately rewarded based on a multiplier to further distil the sticky behaviour regardless of market environments. \\

This is also supported by the excess liquidity paid for by the exchange on normal days and hence, the scheme should be recalibrated to optimize the LP rewards.  As seen in the Xetra Liquidity Provider Programme, an initial constant of 2 is suggested to be multiplied to Q scores on volatile days. \\

\begin{table}[h]
\centering
\caption{Extracted from Xetra Liquidity Programme Agreement}
\begin{tabular}{ccc}
\includegraphics[width=0.8\linewidth]{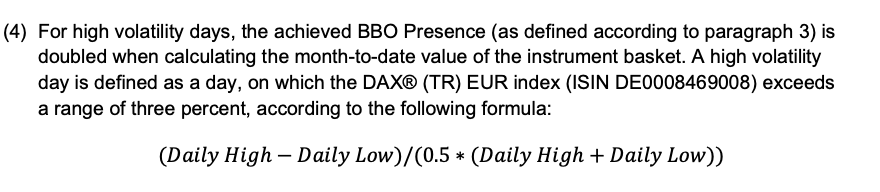} 
\end{tabular}
\end{table}

However, due to the lack of readily available historical data of Q scores for individual markets, the impacts of these should be revisited in the future.

\subsubsection{Implement a liquidity mining campaign with Hummingbot to provide targeted liquidity}

Since the grant for Hummingbot from Q4 2022, this has brought over \$10m of liquidity to dYdX. Despite receiving no publicity, it can be argued that the partnership has achieved moderate success in attracting LPs to the platform. A liquidity mining campaign can be organized to drive liquidity to long tail markets. This partnership has been similarly implemented on other CEXs such as Binance, KuCoin and Gate.io. 

\section{Conclusion}

\par Ultimately, the LP incentives scheme plays a critical role in supporting vibrant markets on dYdX. The parameters have to be carefully calibrated to seek a fine balance between LPs (profitability vs risk) and takers (slippages), which can be further seen from the perspective of the operating spreads and depths permissible. The ideal state of the exchange would be when LPs instinctively compete at the BBO, ensuring a healthy flow of liquidity and this can be effectively witnessed in mature markets (BTC/ETH) given the high volumes, tighter spreads and lower inherent risks. \\

\par Through multiple series of backtesting on historical and more recent data using different metrics, the adjusted maxSpreads have been suggested in Section 6, directing incentives to a  tighter active range in majority of the markets. At the same time, alternatives have been explored to improve the LP rewards mechanism based on the recent end of epoch data. The following are recommended as the immediate steps: 
\par 
1. Revise the LP rewards scheme for BTC/ETH markets to a volume parameter. \\
\par 2. Replace the LP rewards scheme for BTC/ETH markets with the rebates suggested by Wintermute\\
\par 3. Increase the makerVolume weightage for non-BTC/ETH markets from 0.65 to 0.85, following the path taken by BTC/ETH last year. \\
\par 4. Implement Enhanced Rebates for Long Tail / New Markets\\

The impacts on liquidity dynamics will then have to be monitored before adjusting the rewards allocations and introducing a reward multiplier for periods of volatility. The implementation of DMMs for v4 will also entail an interesting discussion for permissionless listings when more information is available. \\

\newpage

\section{References}

\par 1) Anand, A.,  Venkataraman, K. (2010, September 7). Should Exchanges impose Market Maker obligations? \\
 \href{https://www.sec.gov/divisions/riskfin/seminar/venkataraman0313.pdf}{https://www.sec.gov/divisions/riskfin/seminar/venkataraman0313.pdf} \\
\\
2) BIS. (2015, March 18). The economics of market making. The Bank for International Settlements.\\
\href{https://www.bis.org/publ/qtrpdf/r\_qt1503y.html}{https://www.bis.org/publ/qtrpdf/r\_qt1503y.html}   \\
\\
3)  Callen  (2022, April 15). A step towards a more equitable liquidity provider reward structure. Commonwealth.  \\
\href{https://forums.dydx.community/discussion/4407-a-step-towards-a-more-equitable-liquidity-provider-reward-structure} {https://forums.dydx.community/discussion/4407-a-step-towards-a-more-equitable-liquidity-provider-reward-structure}  \\
\\
4) Copeland, T. E.,  Galai, D. (1983). Information effects on the bid--ask spread. The Journal of Finance, 38(5), 1457 –1469.  \\
\href{https://doi.org/10.1111/j.1540-6261.1983.tb03834.x}{https://doi.org/10.1111/j.1540-6261.1983.tb03834.x} \\
\\
5) Demsetz, H. (1968). The cost of transacting. The Quarterly Journal of Economics, 82(1), 33–53.\\
  \href{https://doi.org/10.2307/1882244}{https://doi.org/10.2307/1882244}
\\
\\
6) dYdX. (n.d.). Liquidity Provider Rewards - Governance Documentation. \\
\href{https://docs.dydx.community/dydx-governance/rewards/liquidity-provider-rewards}{https://docs.dydx.community/dydx-governance/rewards/liquidity-provider-rewards}
  \\
\\
7) Lo, D. K., Hall, A. D. (2015, October 9). Resiliency of the Limit Order Book. \\
\href{https://opus.lib.uts.edu.au/bitstream/10453/98964/1/Lo_Hall_Resiliency_of_the_limit_order_book_Accepted_Manuscript.pdf}{https://opus.lib.uts.edu.au/bitstream/10453/98964/1/} \\
\href{Lo_Hall_Resiliency_of_the_limit_order_book_Accepted_Manuscript.pdf}{Lo\_Hall\_Resiliency\_of\_the\_limit\_order\_book\_Accepted\_Manuscript.pdf}
\\
\\
8) Müller (2021, November 13). Incentivize liquidity providers on spread. Commonwealth.\\
  \href{https://forums.dydx.community/discussion/2519-incentivize-liquidity-providers-on-spread} {https://forums.dydx.community/discussion/2519-incentivize-liquidity-providers-on-spread}  \\
\\
9) SLN Capital. (2022, July 20). Discussion- revisions to improve the LP reward formula. Commonwealth.\\
  \href{https://forums.dydx.community/discussion/6322-discussion-revisions-to-improve-the-lp-reward-formula}{https://forums.dydx.community/discussion/6322-discussion-revisions-to-improve-the-lp-reward-formula}  \\
\\

\newpage

\section{Appendix}

\subsection{Min Tick Size vs Spread}
\begin{table}[h]
\centering
\small % Reduce font size
\setlength{\tabcolsep}{2pt} % Adjust column spacing
\begin{tabular}{|l|c|c|c|}
\hline
Market & tickSize & indexPrice & Min Bps \\
\hline
CELO-USD & 0.0010 & 0.5306 & 0.188 \\
LINK-USD & 0.0010 & 6.6375 & 0.015 \\
DOGE-USD & 0.0001 & 0.0725 & 0.138 \\
1INCH-USD & 0.0010 & 0.4171 & 0.240 \\
XMR-USD & 0.1000 & 153.8950 & 0.065 \\
FIL-USD & 0.0100 & 4.5030 & 0.222 \\
ETH-USD & 0.1000 & 1826.0380 & 0.005 \\
AAVE-USD & 0.0100 & 63.6962 & 0.016 \\
ATOM-USD & 0.0010 & 10.9800 & 0.009 \\
MKR-USD & 1.0000 & 637.6600 & 0.157 \\
EOS-USD & 0.0010 & 0.8899 & 0.112 \\
COMP-USD & 0.1000 & 35.2400 & 0.284 \\
ALGO-USD & 0.0001 & 0.1660 & 0.060 \\
XTZ-USD & 0.0010 & 0.8963 & 0.112 \\
UNI-USD & 0.0010 & 5.1756 & 0.019 \\
ADA-USD & 0.0010 & 0.3700 & 0.270 \\
ZRX-USD & 0.0010 & 0.2232 & 0.448 \\
YFI-USD & 1.0000 & 6975.0200 & 0.014 \\
MATIC-USD & 0.0001 & 0.8688 & 0.012 \\
ETC-USD & 0.0100 & 18.3297 & 0.055 \\
AVAX-USD & 0.0100 & 15.2459 & 0.066 \\
LTC-USD & 0.1000 & 86.5360 & 0.116 \\
ENJ-USD & 0.0010 & 0.3414 & 0.293 \\
DOT-USD & 0.0100 & 5.3837 & 0.186 \\
SNX-USD & 0.0010 & 2.1035 & 0.048 \\
RUNE-USD & 0.0010 & 1.1918 & 0.084 \\
XLM-USD & 0.0001 & 0.0884 & 0.113 \\
BCH-USD & 0.1000 & 118.6310 & 0.084 \\
TRX-USD & 0.0001 & 0.0704 & 0.142 \\
BTC-USD & 1.0000 & 27369.0000 & 0.004 \\
UMA-USD & 0.0100 & 2.3360 & 0.428 \\
NEAR-USD & 0.0010 & 1.6745 & 0.060 \\
ZEC-USD & 0.1000 & 33.2750 & 0.301 \\
SOL-USD & 0.0010 & 21.3291 & 0.005 \\
SUSHI-USD & 0.0010 & 0.8999 & 0.111 \\
ICP-USD & 0.0100 & 5.3205 & 0.188 \\
CRV-USD & 0.0001 & 0.8171 & 0.012 \\
\hline
\end{tabular}
\end{table}

\newpage
\begin{landscape}
\subsection{Average Orderbook Depths Recorded}

\begin{table}[!ht]
    \centering
{\tiny
    \begin{tabular}{cccccccccccccc}
    \hline
        ~ & Market & 10bps & -10bps & 15bps & -15bps & 20bps & -20bps & 30bps & -30bps & 40bps & -40bps \\ \hline
        0 & CELO-USD & 28787.73 & 35185.46 & 29261.51 & 35850.78 & 29669.73 & 36290.08 & 191409.78 & 188850.3 & 195142.8 & 192413.53 \\ \hline
        1 & LINK-USD & 161107.56 & 134163.24 & 303082.62 & 286661.05 & 431983.09 & 436506.34 & 609090.46 & 633756.96 & 771597.82 & 817821.25 \\ \hline
        2 & DOGE-USD & 157078.05 & 168368.21 & 157180.16 & 168495.85 & 157588.06 & 168992.95 & 515988.06 & 549654.67 & 738602.04 & 801575.77 \\ \hline
        3 & 1INCH-USD & 0.0 & 0.0 & 109029.13 & 134063.86 & 109029.13 & 134063.86 & 109029.13 & 134063.86 & 576846.92 & 609875.06 \\ \hline
        4 & XMR-USD & 92587.85 & 83628.09 & 114543.19 & 103350.03 & 228441.84 & 221479.12 & 411085.52 & 445251.56 & 548605.38 & 611941.5 \\ \hline
        5 & FIL-USD & 0.0 & 0.0 & 251647.17 & 239890.42 & 251647.17 & 239890.42 & 251725.48 & 240029.29 & 850872.67 & 879431.32 \\ \hline
        6 & ETH-USD & 2546580.58 & 2561040.27 & 4245665.63 & 4397182.39 & 5708734.8 & 5890386.79 & 6599574.08 & 6788933.85 & 7187918.98 & 7426181.35 \\ \hline
        7 & AAVE-USD & 49697.67 & 70134.63 & 130893.5 & 136773.32 & 216724.07 & 213279.09 & 341797.6 & 368186.34 & 638373.2 & 681698.27 \\ \hline
        8 & ATOM-USD & 115717.71 & 110604.11 & 219229.62 & 214303.62 & 297760.16 & 316492.55 & 405596.36 & 443282.39 & 685211.8 & 735087.55 \\ \hline
        9 & MKR-USD & 34586.36 & 32854.05 & 34586.36 & 32854.05 & 34635.74 & 32886.78 & 250517.94 & 247363.69 & 322585.42 & 330185.33 \\ \hline
        10 & EOS-USD & 55997.53 & 51212.27 & 56538.9 & 51618.22 & 290392.09 & 313614.02 & 651776.29 & 670471.04 & 734216.44 & 745643.36 \\ \hline
        11 & COMP-USD & 0.0 & 0.0 & 100175.95 & 128884.87 & 100175.95 & 128884.87 & 100180.43 & 128894.51 & 100180.43 & 128894.51 \\ \hline
        12 & ALGO-USD & 74774.17 & 65591.89 & 105661.85 & 90547.56 & 186312.39 & 164822.37 & 312329.76 & 334672.3 & 562642.95 & 598113.26 \\ \hline
        13 & XTZ-USD & 52945.44 & 38800.14 & 64942.87 & 50372.23 & 265699.24 & 237051.89 & 617995.99 & 630561.02 & 690939.83 & 729702.0 \\ \hline
        14 & UNI-USD & 67701.46 & 65159.16 & 169719.46 & 164895.65 & 259421.98 & 262384.3 & 387069.44 & 427436.78 & 692131.05 & 753413.52 \\ \hline
        15 & ADA-USD & 0.0 & 0.0 & 266608.77 & 249105.21 & 266608.77 & 249105.21 & 266708.14 & 249188.85 & 300039.47 & 278033.12 \\ \hline
        16 & ZRX-USD & 0.0 & 0.0 & 0.0 & 0.0 & 0.0 & 0.0 & 122900.49 & 107576.54 & 122900.49 & 107576.54 \\ \hline
        17 & YFI-USD & 54557.0 & 41524.0 & 118384.17 & 92580.62 & 172234.09 & 150188.31 & 281996.09 & 299596.37 & 540530.75 & 567477.19 \\ \hline
        18 & MATIC-USD & 155602.0 & 153258.56 & 299957.34 & 310325.85 & 459757.23 & 500703.95 & 709877.37 & 768827.86 & 867585.38 & 903852.39 \\ \hline
        19 & ETC-USD & 83555.65 & 80285.83 & 249551.01 & 196636.3 & 350814.05 & 288537.03 & 461066.18 & 424340.71 & 652954.32 & 620185.48 \\ \hline
        20 & AVAX-USD & 71156.85 & 65520.86 & 245219.77 & 209706.55 & 447134.5 & 428839.9 & 582101.62 & 587861.68 & 868630.57 & 918474.46 \\ \hline
        21 & LTC-USD & 104308.24 & 121798.89 & 104483.89 & 122024.11 & 455209.41 & 439667.45 & 769702.73 & 733563.39 & 877667.67 & 821924.51 \\ \hline
        22 & ENJ-USD & 0.0 & 0.0 & 99556.63 & 97717.75 & 115190.17 & 113647.88 & 115227.99 & 113691.18 & 115231.37 & 113696.72 \\ \hline
        23 & DOT-USD & 248586.62 & 183757.21 & 248586.62 & 183757.21 & 248710.7 & 183839.72 & 998962.68 & 744084.09 & 999031.7 & 744151.4 \\ \hline
        24 & SNX-USD & 27141.83 & 25747.3 & 86026.61 & 78815.72 & 125804.03 & 113801.21 & 203463.87 & 190677.65 & 392287.28 & 397621.78 \\ \hline
        25 & RUNE-USD & 13124.4 & 14237.15 & 70953.94 & 74786.44 & 80483.43 & 82964.79 & 235049.24 & 233721.19 & 483292.95 & 511799.25 \\ \hline
        26 & XLM-USD & 66146.3 & 68276.43 & 66280.72 & 68429.97 & 306274.56 & 342618.05 & 684292.57 & 717040.94 & 742069.3 & 791883.39 \\ \hline
        27 & BCH-USD & 48738.17 & 54009.16 & 275395.6 & 281682.72 & 276658.22 & 282958.93 & 527205.42 & 557679.96 & 792609.95 & 853019.14 \\ \hline
        28 & TRX-USD & 108412.16 & 113707.18 & 108427.64 & 113716.17 & 226815.75 & 228915.3 & 520635.16 & 478520.6 & 857417.44 & 793290.23 \\ \hline
        29 & BTC-USD & 2276943.17 & 2174704.49 & 3510947.24 & 3607486.39 & 4567733.64 & 4749037.02 & 5275372.07 & 5471668.8 & 5844803.31 & 5974759.74 \\ \hline
        30 & UMA-USD & 0.0 & 0.0 & 0.0 & 0.0 & 0.0 & 0.0 & 34135.46 & 37805.43 & 34135.46 & 37805.43 \\ \hline
        31 & NEAR-USD & 121450.82 & 113877.07 & 171431.81 & 170266.38 & 311177.15 & 313344.63 & 502781.99 & 544324.1 & 662886.96 & 713059.54 \\ \hline
        32 & ZEC-USD & 0.0 & 0.0 & 44215.3 & 48066.66 & 137469.38 & 155516.76 & 137469.38 & 155516.76 & 137469.38 & 155516.76 \\ \hline
        33 & SOL-USD & 185227.36 & 164849.59 & 363538.86 & 333888.11 & 519990.69 & 477412.04 & 739261.33 & 728070.08 & 896721.74 & 899660.61 \\ \hline
        34 & SUSHI-USD & 20361.69 & 19384.7 & 21399.07 & 20517.44 & 108619.89 & 97515.15 & 407091.82 & 419850.96 & 459266.71 & 484903.72 \\ \hline
        35 & ICP-USD & 104987.07 & 102524.32 & 124937.43 & 121535.02 & 125132.25 & 121722.91 & 501373.47 & 552225.15 & 577716.97 & 636908.5 \\ \hline
        36 & CRV-USD & 43927.14 & 28278.95 & 94219.46 & 77564.36 & 154005.77 & 155217.57 & 268398.71 & 306948.46 & 540221.98 & 591512.26 \\ \hline
    \end{tabular}
    }
\end{table}

\end{landscape}
\newpage

\begin{landscape}

\subsection{Average Daily Trade Statistics - Between April 30 and May 19 2023}

\begin{table}[h]
\centering

\tiny % Reduce font size
\setlength{\tabcolsep}{1pt} % Adjust column spacing
{\tiny
\begin{tabular}{ccccccccc}
    \hline
        Market & avg\_depth\_bid & avg\_depth\_ask & median\_depth\_bid & median\_depth\_ask & max\_depth\_bid & max\_depth\_ask & ninetyfive\_depth\_ask & ninetyfive\_depth\_bid \\ \hline
        1INCH-USD & 2789.27 & 2755.95 & 893.45 & 931.68 & 93212.9 & 96751.7 & 9527.51 & 9160.68 \\ \hline
        AAVE-USD & 1893.59 & 1814.02 & 286.9 & 299.58 & 118885.34 & 94218.78 & 8007.21 & 7894.17 \\ \hline
        ADA-USD & 4166.26 & 4056.73 & 1041.31 & 942.8 & 158795.53 & 182847.56 & 15050.15 & 15778.71 \\ \hline
        ALGO-USD & 1790.31 & 1833.26 & 191.71 & 395.51 & 101664.74 & 78072.64 & 7312.07 & 7004.38 \\ \hline
        ATOM-USD & 2437.53 & 2319.11 & 894.63 & 899.81 & 235158.65 & 190300.3 & 7237.29 & 7012.76 \\ \hline
        AVAX-USD & 3034.81 & 2802.44 & 310.16 & 315.11 & 287715.82 & 248930.04 & 10801.78 & 10671.71 \\ \hline
        BCH-USD & 2371.36 & 2387.52 & 1066.53 & 1002.29 & 124514.71 & 124993.46 & 7882.95 & 8067.13 \\ \hline
        CELO-USD & 1852.56 & 1761.57 & 320.74 & 289.63 & 86065.32 & 72867.64 & 7038.69 & 7180.7 \\ \hline
        COMP-USD & 3682.5 & 3702.15 & 956.42 & 1115.38 & 128628.5 & 110899.83 & 14876.44 & 15285.51 \\ \hline
        CRV-USD & 2552.19 & 2385.84 & 727.5 & 693.15 & 145151.7 & 127082.33 & 9255.79 & 9584.03 \\ \hline
        DOGE-USD & 4192.52 & 4011.55 & 221.07 & 128.25 & 248201.33 & 201763.27 & 19024.23 & 19214.49 \\ \hline
        DOT-USD & 2406.03 & 2341.25 & 519.28 & 482.92 & 156940.46 & 147332.58 & 7979.97 & 8443.82 \\ \hline
        ENJ-USD & 2805.47 & 2793.77 & 180.27 & 192.81 & 110451.14 & 92532.19 & 13127.74 & 12650.09 \\ \hline
        EOS-USD & 2178.94 & 1965.5 & 648.4 & 563.09 & 134356.3 & 92423.47 & 7677.63 & 7771.4 \\ \hline
        ETC-USD & 1397.04 & 1366.84 & 527.37 & 527.47 & 85268.71 & 97027.91 & 3889.54 & 3877.27 \\ \hline
        FIL-USD & 3812.48 & 3626.75 & 870.18 & 867.42 & 188564.73 & 147028.17 & 13623.9 & 13411.81 \\ \hline
        ICP-USD & 3129.93 & 2718.67 & 1294.17 & 918.09 & 160428.42 & 154460.42 & 8831.99 & 10070.65 \\ \hline
        LINK-USD & 2079.48 & 2096.95 & 505.05 & 497.53 & 136568.44 & 134693.74 & 8414.11 & 8144.7 \\ \hline
        LTC-USD & 3060.99 & 3009.72 & 788.42 & 714.4 & 315470.91 & 274313.07 & 8926.51 & 8711.25 \\ \hline
        MATIC-USD & 4966.33 & 5150.44 & 761.81 & 801.63 & 335618.61 & 339231.69 & 20824.43 & 19964.91 \\ \hline
        MKR-USD & 3057.52 & 3114.05 & 696.84 & 773.58 & 166510.62 & 158854.52 & 11223.02 & 9785.69 \\ \hline
        NEAR-USD & 2215.09 & 2174.38 & 412.18 & 416.58 & 142771.98 & 128762.89 & 7666.1 & 7299.51 \\ \hline
        RUNE-USD & 2067.09 & 2088.74 & 380.08 & 396.84 & 71298.52 & 92186.51 & 7256.69 & 7294.08 \\ \hline
        SNX-USD & 1370.81 & 1314 & 301.61 & 294.16 & 98496.77 & 71231.53 & 5312.83 & 5186.17 \\ \hline
        SOL-USD & 9823.7 & 8276.23 & 2806.47 & 2542.62 & 677403.74 & 434800 & 31304.66 & 32501.9 \\ \hline
        SUSHI-USD & 2625.56 & 2548.38 & 577.01 & 526.65 & 150481.78 & 122771.27 & 10003.4 & 9857.84 \\ \hline
        TRX-USD & 1576.53 & 1748.46 & 8.85 & 15.85 & 58391.89 & 74238.7 & 7623.28 & 6756.08 \\ \hline
        UMA-USD & 2805.17 & 2696.41 & 583.65 & 430.85 & 54885.67 & 47593.37 & 11875.47 & 11797.12 \\ \hline
        UNI-USD & 1835.82 & 1782.74 & 514.12 & 527.07 & 100119.74 & 91063.31 & 6371.7 & 6258.08 \\ \hline
        XLM-USD & 1414.41 & 1436.1 & 109.73 & 137.59 & 90328.01 & 75175.22 & 5474.3 & 5029.96 \\ \hline
        XMR-USD & 2001.54 & 2012.58 & 543.64 & 456.39 & 71020.94 & 74122.01 & 7350.28 & 7008.55 \\ \hline
        XTZ-USD & 2870.74 & 3167.37 & 613.34 & 1273.41 & 98378.35 & 96117.9 & 12174.94 & 11342.99 \\ \hline
        YFI-USD & 2027.92 & 1984.43 & 571.36 & 557.04 & 58333.59 & 59568.33 & 7148.84 & 7249.9 \\ \hline
        ZEC-USD & 2783.02 & 2742.66 & 293.72 & 326.47 & 88542.51 & 72485.04 & 11941.19 & 12102.72 \\ \hline
        ZRX-USD & 1903.7 & 1869.52 & 4.08 & 2.79 & 48038.94 & 40706.58 & 10648.94 & 10867.31 \\ \hline
    \end{tabular}}
\end{table}
\end{landscape}
\newpage

\begin{landscape}

\subsection{Average Daily Trade Statistics - Between May 20 and May 24 2023}

\begin{table}[h]
\centering

\tiny % Reduce font size
\setlength{\tabcolsep}{1pt} % Adjust column spacing
{\tiny
\begin{tabular}{ccccccccc}
    \hline
        Market & avg\_depth\_bid & avg\_depth\_ask & median\_depth\_bid & median\_depth\_ask & max\_depth\_bid & max\_depth\_ask & ninetyfive\_depth\_ask & ninetyfive\_depth\_bid \\ \hline
        1INCH-USD & 2290.579044 & 2460.290254 & 361.4916 & 563.6371 & 109632.3542 & 121802.8124 & 7198.18136 & 6888.13844 \\ \hline
        AAVE-USD & 775.6549121 & 761.5169433 & 51.07588 & 58.70494 & 73446.60082 & 69163.46306 & 3757.164317 & 4136.08648 \\ \hline
        ADA-USD & 3820.347317 & 3857.431758 & 357.3074 & 251.37 & 127320.7402 & 111137.0242 & 15085.7699 & 14702.23388 \\ \hline
        ALGO-USD & 922.6858346 & 984.4883105 & 4.93128 & 97.93578 & 70966.42644 & 53629.45504 & 3538.859346 & 3389.38457 \\ \hline
        ATOM-USD & 1121.430254 & 1064.538328 & 589.58854 & 586.63624 & 80743.50236 & 67970.915 & 2984.20342 & 3054.897074 \\ \hline
        AVAX-USD & 2177.6453 & 2176.558451 & 298.0914 & 293.0652 & 172432.2334 & 147763.7586 & 8250.7979 & 7662.60364 \\ \hline
        BCH-USD & 1845.875327 & 1866.546088 & 248.158 & 191.2816 & 121276.1044 & 130084.8248 & 7101.93904 & 7143.54892 \\ \hline
        BTC-USD & 56450.69792 & 57942.51278 & 9511.76117 & 10006.6117 & 2830382.275 & 2180924.896 & 240256.2621 & 234561.5017 \\ \hline
        CELO-USD & 1637.066263 & 1563.510474 & 209.1639 & 161.5299 & 84141.8776 & 62433.8058 & 6429.0748 & 6672.86464 \\ \hline
        COMP-USD & 1930.978384 & 2277.769704 & 24.3419 & 102.9572 & 52415.2262 & 71506.9772 & 10811.32928 & 10575.63307 \\ \hline
        CRV-USD & 1558.056675 & 1439.200968 & 391.18707 & 358.74766 & 123526.1518 & 77974.43934 & 5184.467106 & 4880.274604 \\ \hline
        DOGE-USD & 3332.401806 & 3567.755459 & 70.8678 & 87.2766 & 162941.9924 & 142367.026 & 19198.87974 & 16649.4005 \\ \hline
        DOT-USD & 1937.306672 & 1924.97545 & 96.1776 & 17.4346 & 148426.7554 & 128115.0674 & 7946.63104 & 7497.7291 \\ \hline
        ENJ-USD & 1629.321883 & 1680.924461 & 206.7679 & 189.5885 & 44442.1874 & 34013.5104 & 8639.21774 & 8123.93421 \\ \hline
        EOS-USD & 1615.56714 & 1489.637366 & 168.0852 & 173.325 & 112854.525 & 107807.2018 & 5214.70379 & 6477.58528 \\ \hline
        ETC-USD & 821.335813 & 834.5595786 & 269.6208 & 355.525 & 97193.5846 & 89041.3986 & 2400.02366 & 2506.95224 \\ \hline
        ETH-USD & 72858.82878 & 76748.17239 & 13473.43308 & 13180.07441 & 3367337.936 & 2951386.36 & 347178.9095 & 332066.827 \\ \hline
        FIL-USD & 2845.570969 & 2375.42143 & 242.1354 & 306.0274 & 283558.3276 & 178139.4616 & 6862.59637 & 7448.04701 \\ \hline
        ICP-USD & 1495.919987 & 1380.981044 & 140.767 & 235.812 & 138953.3468 & 75877.6646 & 4459.65027 & 4107.79114 \\ \hline
        LINK-USD & 1420.933924 & 1464.899258 & 278.73686 & 197.37878 & 121695.0593 & 87505.56024 & 5385.099861 & 4959.656777 \\ \hline
        LTC-USD & 2662.874788 & 2607.647309 & 529.7208 & 491.9911 & 243949.2008 & 281985.1228 & 8204.50186 & 8422.61403 \\ \hline
        MATIC-USD & 2536.885646 & 2830.372519 & 343.54898 & 344.19116 & 137469.4662 & 316325.6192 & 10919.27501 & 10208.46728 \\ \hline
        MKR-USD & 1750.639831 & 1732.809627 & 198.5781 & 35.5703 & 117634.958 & 155068.8478 & 4622.26376 & 4199.58468 \\ \hline
        NEAR-USD & 1675.165519 & 1562.516707 & 253.3468 & 165.22656 & 88399.77032 & 95291.04754 & 5112.051726 & 5469.60348 \\ \hline
        RUNE-USD & 1405.301793 & 1236.978801 & 16.3594 & 14.192 & 95479.7172 & 35884.6144 & 4874.90012 & 5169.0742 \\ \hline
        SNX-USD & 1176.601377 & 1181.77794 & 175.40392 & 157.22626 & 74976.70678 & 63317.19966 & 4317.679256 & 4539.195001 \\ \hline
        SOL-USD & 5593.702429 & 4965.523614 & 598.20531 & 553.09842 & 510629.164 & 325720.7934 & 19811.2121 & 17715.71406 \\ \hline
        SUSHI-USD & 1742.433093 & 1702.580362 & 451.14212 & 363.43854 & 119696.5731 & 97862.32946 & 6188.750554 & 5853.849843 \\ \hline
        TRX-USD & 2020.263181 & 2206.362237 & 207.119 & 186.352 & 112130.8632 & 100583.9748 & 8609.58809 & 7762.54352 \\ \hline
        UMA-USD & 1674.146218 & 1867.435641 & 1.661 & 2.3113 & 27597.5654 & 25958.7702 & 10339.02843 & 9421.88208 \\ \hline
        UNI-USD & 1228.737573 & 1290.555566 & 19.93088 & 110.55462 & 60221.73284 & 79114.6061 & 5118.353054 & 4848.283492 \\ \hline
        XLM-USD & 945.8556346 & 1056.661455 & 14.4154 & 11.2268 & 27968.5548 & 33075.7324 & 4160.83337 & 3593.48083 \\ \hline
        XMR-USD & 1241.302452 & 1311.831047 & 149.1158 & 284.344 & 60161.3878 & 47362.076 & 4184.67491 & 3944.77843 \\ \hline
        XTZ-USD & 2535.35591 & 2924.513667 & 25.3692 & 654.9378 & 118821.1654 & 116389.5882 & 12495.11196 & 11254.26466 \\ \hline
        YFI-USD & 1141.032228 & 1067.767459 & 30.76762 & 74.41496 & 57176.42934 & 51629.6775 & 4650.909977 & 4621.203557 \\ \hline
        ZEC-USD & 1498.67572 & 1417.133854 & 3.5894 & 4.6175 & 46219.6362 & 43259.5068 & 7728.24865 & 7725.8727 \\ \hline
        ZRX-USD & 1095.318171 & 1103.402167 & 1.8322 & 3.1085 & 17729.5864 & 18748.5542 & 7004.15086 & 6953.47491 \\ \hline
    \end{tabular}}
\end{table}

\end{landscape}
\newpage

\subsection{PPI Release with Tighter Spreads}

\includegraphics[width=0.8\textwidth]{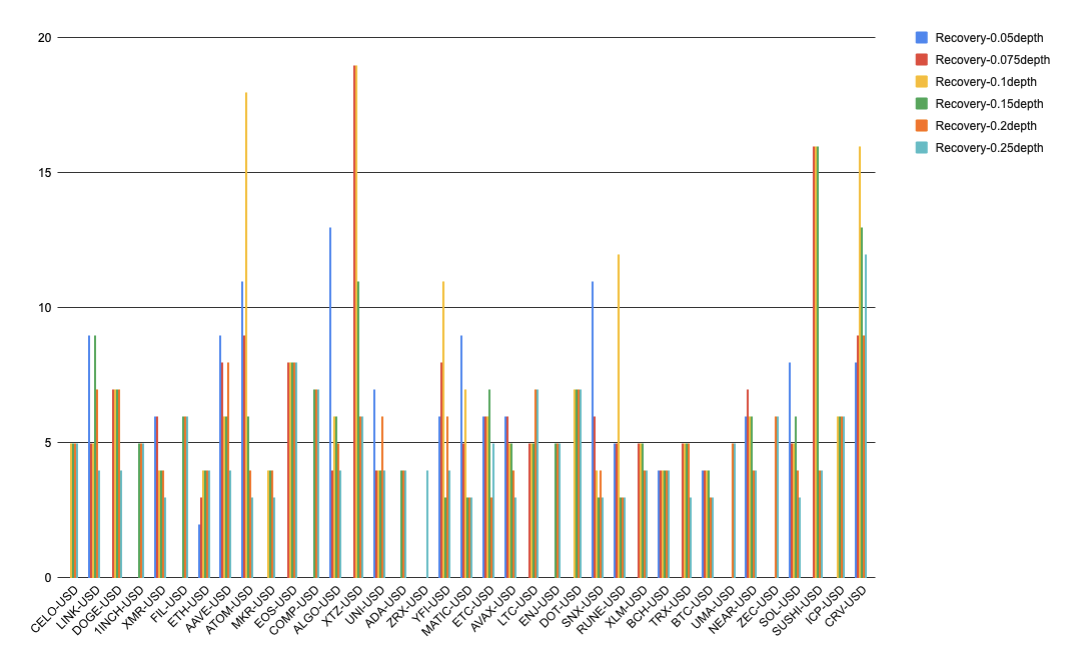}\\
\includegraphics[width=0.8\textwidth]{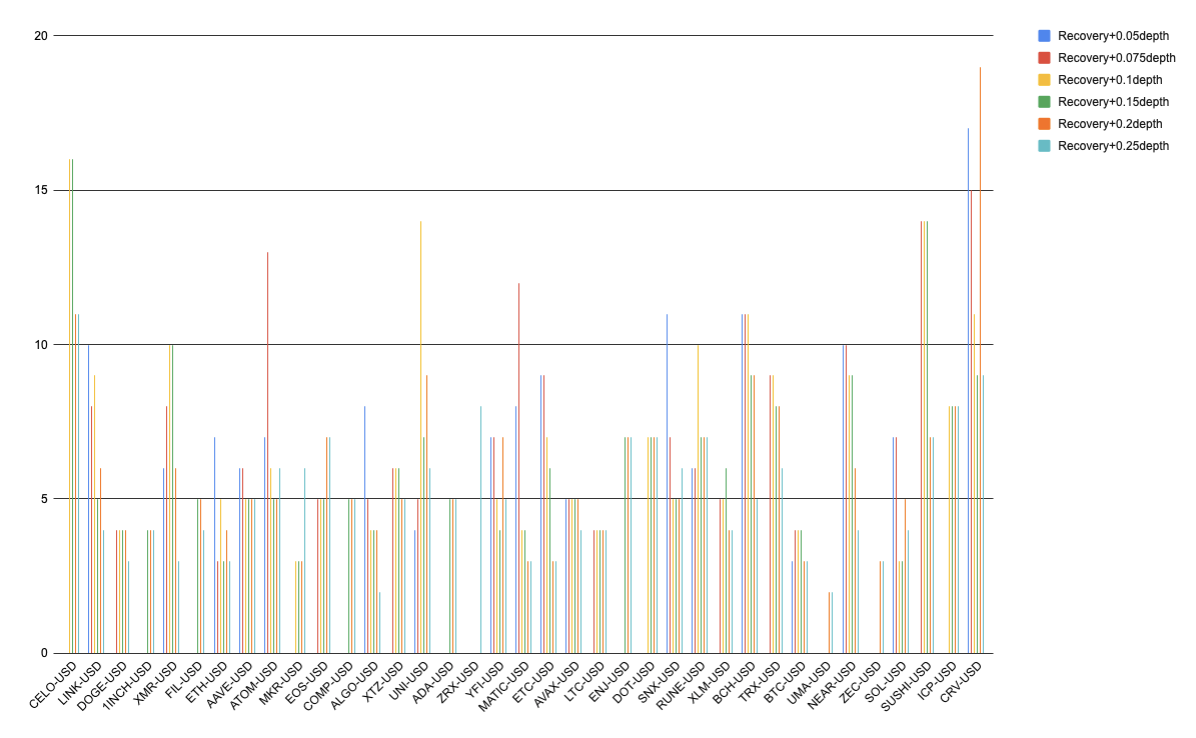}

\subsection{US Initial Jobless Claims}

\includegraphics[width=0.8\textwidth]{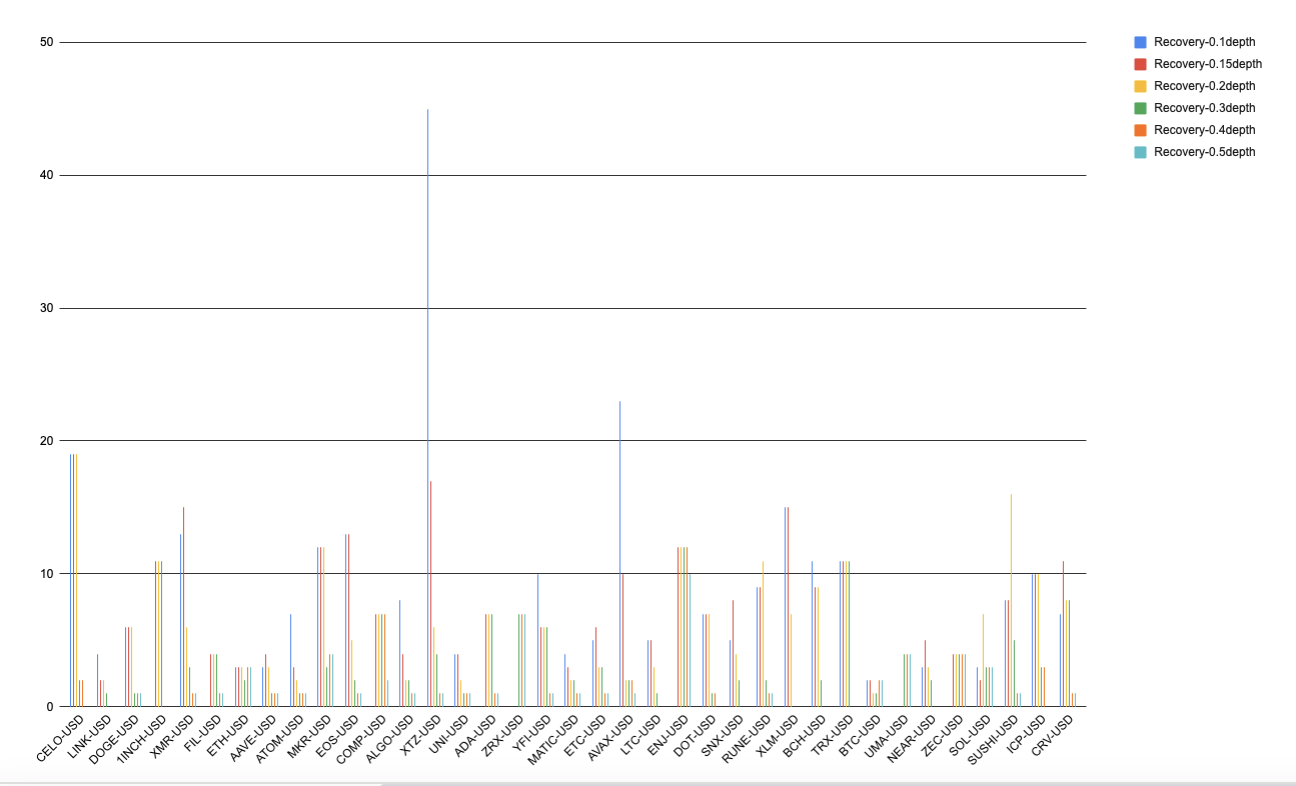}\\
\includegraphics[width=0.8\textwidth]{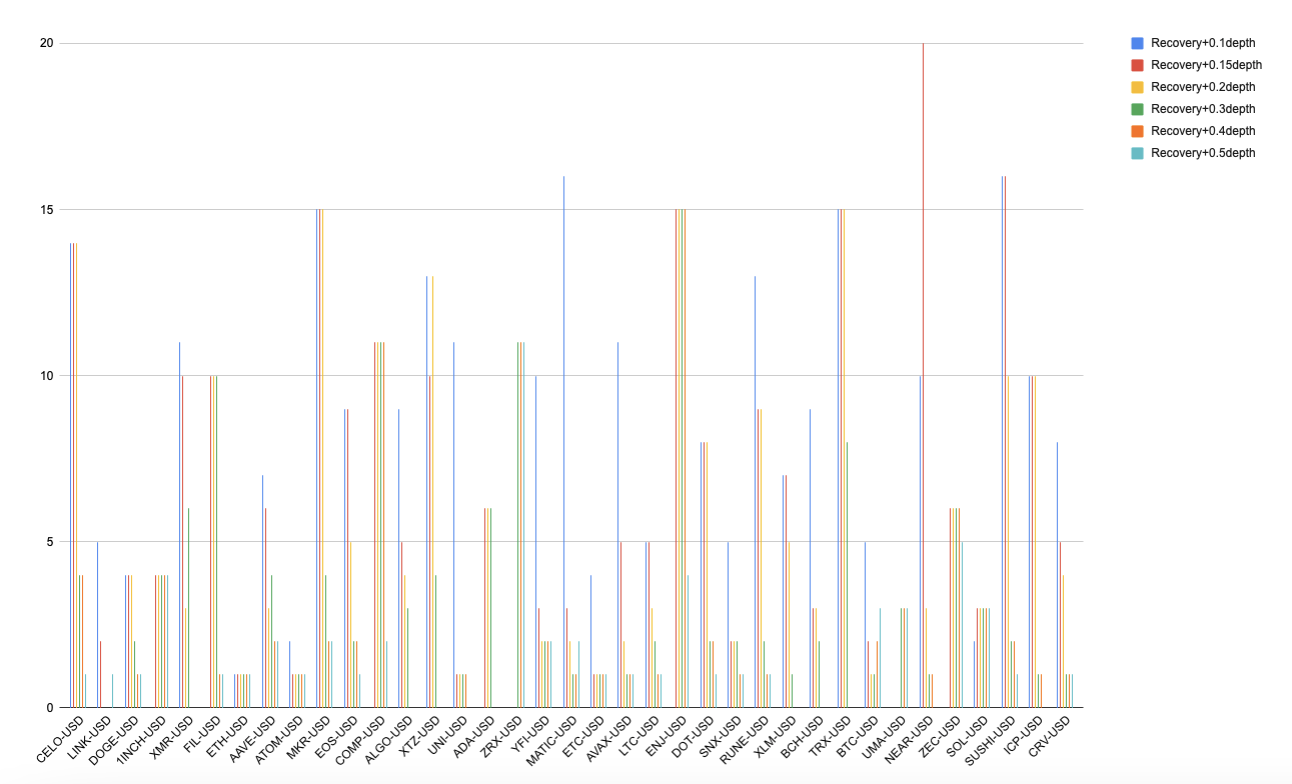}

\newpage
\subsection{FOMC Meeting Minutes Release}

\includegraphics[width=0.8\textwidth]{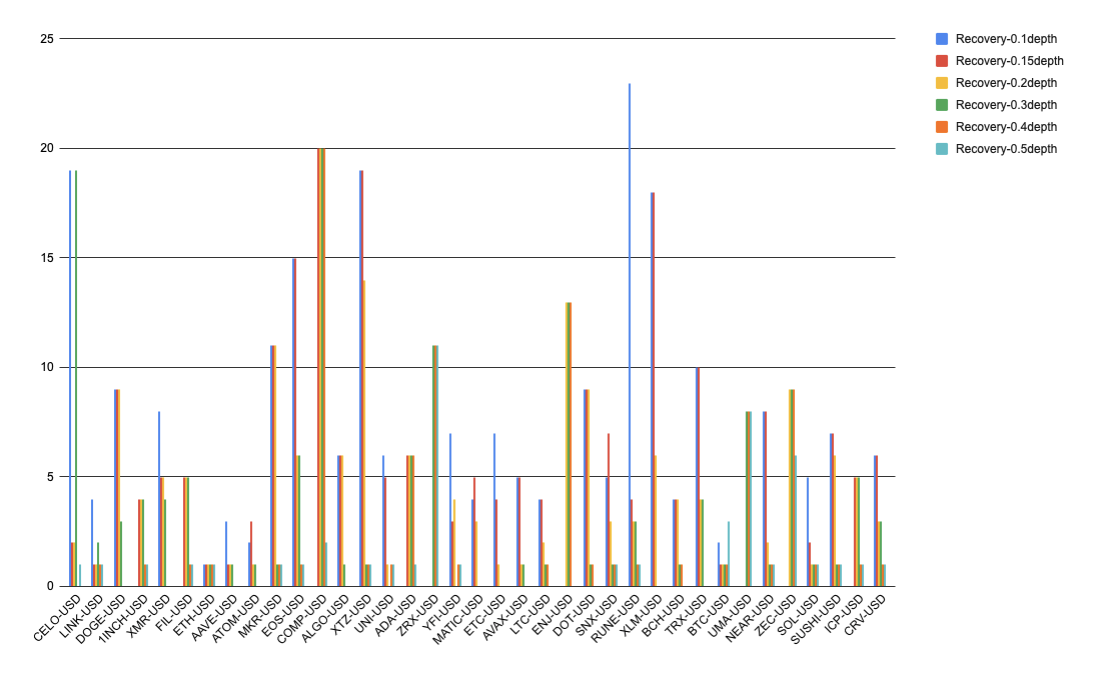}\\
\includegraphics[width=0.8\textwidth]{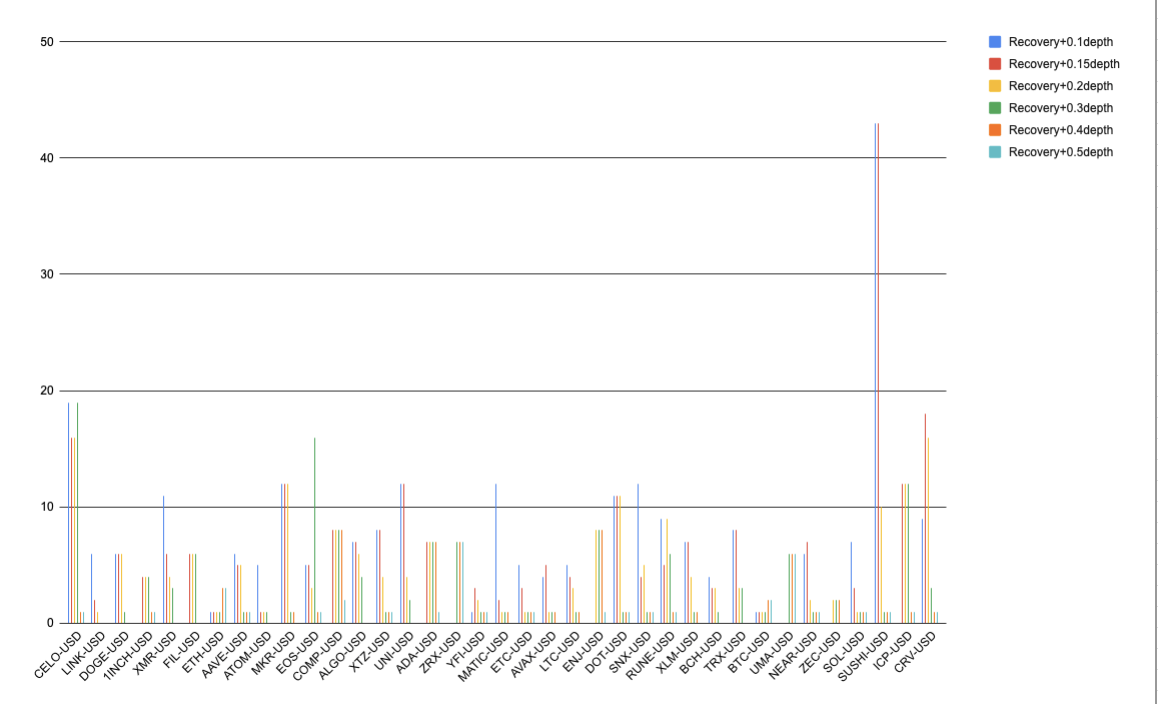}

\newpage
\subsection{Fees Generated}

\begin{table}[!ht]
    \centering
    \begin{tabular}{|l|l|}
    \hline
        Market & Fees \\ \hline
        UMA-USD & 5587.050283 \\ \hline
        RUNE-USD & 11309.512089 \\ \hline
        ZEC-USD & 11740.140657 \\ \hline
        ZRX-USD & 12358.879347 \\ \hline
        ENJ-USD & 13255.440886 \\ \hline
        CELO-USD & 13408.904137 \\ \hline
        COMP-USD & 14309.80742 \\ \hline
        XLM-USD & 15333.464136 \\ \hline
        ETC-USD & 15570.710288 \\ \hline
        UNI-USD & 16431.278735 \\ \hline
        YFI-USD & 16512.778395 \\ \hline
        XMR-USD & 16620.451919 \\ \hline
        MKR-USD & 17944.658857 \\ \hline
        BCH-USD & 18684.292511 \\ \hline
        XTZ-USD & 19308.532013 \\ \hline
        1INCH-USD & 19428.0062 \\ \hline
        ALGO-USD & 19914.182586 \\ \hline
        ICP-USD & 22817.903744 \\ \hline
        DOT-USD & 23420.763145 \\ \hline
        EOS-USD & 23538.252361 \\ \hline
        NEAR-USD & 23540.184115 \\ \hline
        TRX-USD & 24342.576194 \\ \hline
        AAVE-USD & 24537.743053 \\ \hline
        SNX-USD & 25241.952879 \\ \hline
        LINK-USD & 31594.305213 \\ \hline
        SUSHI-USD & 32487.182111 \\ \hline
        ATOM-USD & 32988.331918 \\ \hline
        FIL-USD & 40869.382344 \\ \hline
        CRV-USD & 43075.945467 \\ \hline
        DOGE-USD & 43079.828927 \\ \hline
        ADA-USD & 47467.200754 \\ \hline
        AVAX-USD & 52823.149907 \\ \hline
        LTC-USD & 67774.840313 \\ \hline
        MATIC-USD & 128119.256386 \\ \hline
        SOL-USD & 183242.177007 \\ \hline
        BTC-USD & 1527581.557764 \\ \hline
        ETH-USD & 1944971.095694 \\ \hline
    \end{tabular}
\end{table}
\end{document}